\documentclass[12pt,preprint]{aastex}

\newcommand{\bq}{\begin{equation}}
\newcommand{\eq}{\end{equation}}

\newcommand{\ffam}{\hbox{$\,.\!\!^{\prime}$}}
\newcommand{\ffas}{\hbox{$\,.\!\!^{\prime\prime}$}}

\shorttitle{Massive evolved galaxies at z$\ga$5}
\begin{document}

\title{A Population of Massive and Evolved Galaxies at z$\ga$5}
\author{T. Wiklind\altaffilmark{1,2},
M. Dickinson\altaffilmark{3},
H. C. Ferguson\altaffilmark{1},
M. Giavalisco\altaffilmark{1},
B. Mobasher\altaffilmark{1,2},
N. A. Grogin\altaffilmark{4},
N. Panagia\altaffilmark{1}
}
\altaffiltext{1}{Space Telescope Science Institute,
3700 San Martin Drive, Baltimore MD 21218, USA; wiklind@stsci.edu}
\altaffiltext{2}{Affiliated with the Space Sciences Department of the European Space Agency}
\altaffiltext{3}{NOAO, 950 N. Cherry Ave. P.O. 26732, Tucson, AZ 85726-6732, USA}
\altaffiltext{4}{School of Earth and Space Exploration, Arizona State University,
P.O. Box 871404, Tempe, AZ 85287-1404, USA}

\begin{abstract}
We report results from a search for massive and evolved galaxies at $z \ga 5$
in the Great Observatories Origins Deep Survey (GOODS) southern field. 
Combining HST ACS, VLT ISAAC and Spitzer IRAC broad--band photometric data, we
develop a color selection technique to identify candidates for being evolved galaxies at
high redshifts. The color selection is primarily based on locating the Balmer--break
using the K- and 3.6$\mu$m bands. Stellar population synthesis models are fitted
to the SEDs of these galaxies to identify the final sample. We find 11 candidates
with photometric redshifts in the range $4.9 \leq z < 6.5$, dominated by an old stellar
population, with ages 0.2$-$1.0 Gyr, and stellar masses in the range $(0.5 - 5) \times
10^{11}$ M$_{\odot}$.
The majority of the stars in these galaxies were formed at $z > 9$ and the current
star formation activity is in all cases, except two, a few percent of the inferred
early star formation rate. One candidate has a spectroscopically confirmed redshift,
in good agreement with our photometric redshift. The galaxies are very compact,
with half--light radii in the observed $K-$band smaller than $\sim 2$ kpc.
Seven of the 11 candidates are also detected at 24$\mu$m with the MIPS instrument
on Spitzer.
%%%
By itself, the 24$\mu$m emission could potentially be interpreted as PAH emission
from a dusty starburst at $z \sim 2-3$, however, it is also consistent with the
presence of an obscured AGN at $z \ga 5$. Indeed, for the $z \ga 5$ solutions,
all the MIPS detected galaxies, except two,
%BBG\#3348 and JD2,
have relatively high
internal extinction. While we favor the obscured AGN interpretation, based on the
model SED fits to the optical/UV, we define a 'no--MIPS' sample of candidates in
addition to the full sample. Results will be quoted for both samples.
We estimate the completeness of the Balmer break galaxy sample to be $\sim$40\%
(an upper limit). The comoving number density of galaxies with a stellar mass $\ga 10^{11}$
M$_{\odot}$, at an average redshift $\bar{z} = 5.2$, is $3.9 \times 10^{-5}$ Mpc$^{-3}$
(no--MIPS sample: $1.4 \times 10^{-5}$ Mpc$^{-3}$).
The corresponding stellar mass density is $8 \times 10^{6}$ M$_{\odot}$ Mpc$^{-3}$
(no--MIPS sample: $6.2 \times 10^{6}$ M$_{\odot}$ Mpc$^{-3}$).
The estimated stellar mass density at $\bar{z} = 5.2$ is $2-3$\% of the present
day total stellar mass density and $20-25$\% of the stellar mass density at
$z \sim 2$. If the stellar mass estimates are correct, the presence of these massive
and evolved galaxies when the universe was $\sim$1 Gyr old could suggest that
conversion of baryons into stars proceeded more efficiently in the early universe
than it does today.
\end{abstract}

\keywords{cosmology: observations --- galaxies: formation --- galaxies: high redshift --
galaxies: photometry --- galaxies: evolution}

\section{Introduction}\label{intro}

An important goal of observational cosmology is to understand how stars
are assembled into galaxies and how this is related to the evolution of dark
matter halos. In prevailing hierarchical models, star formation starts out
in low mass systems, which build more massive galaxies
through sequential merging (e.g. White \& Rees 1978; Somerville 2004).
In this picture, the most massive galaxies are found at relatively low
redshifts. Recently, a significant population of galaxies with stellar mass
$\sim 10^{11}$ M$_{\odot}$ has been found at $z \sim 2 - 3$
(cf. Franx et al. 2003; Glazebrook et al. 2004; Fontana et al. 2004; Yan et al.
2004; Daddi et al. 2005a; Rudnick et al. 2006; van Dokkum et al. 2006).
Stellar population synthesis models combined with broad-band photometric data show
that many of these galaxies contain an old stellar population, with ages indicating a
star formation phase within 1$-$3 Gyr after the Big Bang. Moreover,
a number of submillimeter detected galaxies at $z \sim 2 - 3$, which are known
to be massive systems, based on their inferred molecular gas and dynamical mass
estimates (cf. Greve et al. 2005), also appear to contain an old stellar population
with mass $\sim 10^{11}$ M$_{\odot}$ (Borys et al. 2005).
Therefore, a consensus seems to be emerging, that the most massive galaxies seen
today, formed the bulk of their stars within the first $\sim$3 Gyr of cosmic
history (cf. Cimatti et al. 2004; Daddi et al. 2005b; Juneau et al. 2005).
However, it is not known how these stars were assembled into their present host
galaxies, whether this was done during multiple merger events, as proposed in
hierarchical models, or if the stars and their host galaxy are co-eval.
In view of the early formation epoch implied for many of these massive galaxies,
the question whether the formation is hierarchical or monolithic becomes a matter
of semantics as the merger time scale becomes comparable to the dynamical time scale.

Recent ultra--deep surveys, done at wavelengths stretching from the UV to mid--infrared,
have resulted in detection of galaxies and AGNs at even higher redshifts, reaching
into the era of re-ionization. One example is HUDF--JD2, in the Hubble Ultra Deep Field,
which Mobasher et al. (2005) identify as a candidate for a massive, evolved galaxy at
$z=6.5$. The age of this galaxy is estimated to be $\ga$600 Myrs, with a stellar mass
of $\sim 6 \times 10^{11}$ M$_{\odot}$, much larger than the stellar mass of the Milky Way.
The implied age of this galaxy means that  the bulk of the stars were formed
on a short time scale just a few hundred million years after the recombination era.
Other recent studies have used data from the Spitzer Space Telescope to analyze the
stellar masses and ages for galaxies at $z > 5$ (cf. Yan et al. 2005, 2006; Eyles et
al. 2005, 2006; Stark et al. 2006; Verma et al. 2007).
The inferred stellar masses range from $1 - 10 \times 10^{10}$ M$_{\odot}$ and ages
of several $10^8$ years. In several cases, galaxies have spectroscopically determined
redshifts.
Another spectroscopically confirmed galaxy is the gravitationally lensed object HCM06
at $z = 6.56$ (Hu et al. 2002), with a stellar mass of a few $10^{10}$ M$_{\odot}$ and
an age of $\sim$300 Myr (cf. Chary, Stern \& Eisenhardt 2005; Schaerer \& Pell\'{o} 2005). 

The presence of these massive and old galaxies at $z \ga 5$ holds important clues
for understanding how the first galaxies formed and how the galaxy population
in general has evolved with cosmic time.
In order to determine whether a significant population of massive and old galaxies
exists at $z > 5$, and to derive the parameters characterizing this population, we
need a selection method that specifically targets and selects evolved stellar systems
at very high redshifts, using broad-band photometric data available from deep
multiwavelength surveys.
The presence of old galaxies at high redshift
can not efficiently be inferred using the normal Lyman drop-out technique.
The drop-out technique has proven to be efficient in detecting galaxies that
are actively forming stars and contain relatively small amounts of dust, but
it is ineffective for detecting galaxies without strong UV continuum.

In this paper we develop a method for selecting galaxies dominated by a stellar population
older than $\sim$100 Myr and situated at $z \ga 5$, and discuss the results and its implications.
The technique is primarily based on detecting the presence of a well--developed Balmer break,
redshifted to $\sim$3$\mu$m,  that can be probed by the K$_\mathrm{s} - 3.6\mu$m color index.
A second color index is used to further isolate the old high-z galaxies from foreground `contaminants'. 
The color signature of the Balmer break has previously been used to select galaxies at redshifts
$z \sim 1 - 3$ (Franx et al. 2003; Daddi et al. 2005a; Adelberger et al. 2004).
By choosing a suitable filter combination, the Balmer break can be used to select
galaxies at any redshift, in a manner similar to the Lyman--break technique.
In this paper we will refer to the galaxies selected through this technique at $z > 5$ as
Balmer--break galaxies (BBGs).

The paper is structured as follows: 
In Sect~\ref{data} we present our sample and photometric data.
Sect.~\ref{selecting} discusses the Balmer break feature, the stellar population
synthesis models used and examines the confidence of the model fitting procedure
using Monte Carlo simulations. In this section we also discuss degeneracies and define
the final color selection criteria used in this paper.
In Sect.~\ref{results} we present the Balmer--break candidates selected using our color
criteria and model fits of synthetic stellar spectra. We derive associated physical
parameters from the models and discuss the Spitzer/MIPS 24$\mu$m detections. In this
section we also discuss the individual sources and assign a confidence classification
to each source based on its likelihood to have the correct redshift.
In Sect.~\ref{testing} we apply our model fitting to galaxies with known spectroscopic
redshift and assess the reliability of the estimated parameters.
In Sect.~\ref{errors} we discuss different sources of errors and derive the
completeness of our sample.
We discuss our results in Sect.~\ref{discussion} and compare the number density of
Balmer--break galaxies with the expected number density of dark matter halos.
Sect.~\ref{summary} gives a summary of our results.
We adopt H$_0 = 72$ km s$^{-1}$ Mpc$^{-1}$, $\Omega_m = 0.3$, and $\Omega_{\Lambda} = 0.7$
throughout this paper. All magnitudes are in the AB system (Oke 1974).

\section{The Catalog and Photometric Data}\label{data}

The sample used in this study is selected from the Great Observatories
Origins Deep Survey (GOODS) southern field (Dickinson \& Giavalisco 2003).
This field has been observed at many wavelengths, including optical
(HST/ACS- BV{\it iz}) -- (Giavalisco et al. 2004), near--infrared
(VLT/ISAAC- $JHK_s$) -- (Vandame et al., in prep.), and deep mid--infrared
imaging with the Spitzer Space Telescope with IRAC 
(3.6, 4.5, 5.7 and 8.0$\mu$m) -- (Dickinson et al., in prep.) and 
MIPS (24$\mu$m) -- (Dickinson et al., in prep.) instruments.

The HST-ACS images were obtained in four bands: F435W ($B_{435}$),
F606W ($V_{606}$), F775W ($i_{775}$) and F850LP ($z_{850}$) to limiting
sensitivities of 27.8, 27.8, 27.1 and 26.1 AB magnitudes (10$\sigma$
for an extended source measured over 0.2 arcsec$^2$ aperture) respectively.
We use the ESO v1.5 public release of the GOODS--S ISAAC images. ÊThese
cover 156 arcmin$^2$ in Êthe $J$ and $K_s$ bands, and a somewhat smaller
region, 124~arcmin$^2$, in the $H$ band.
The ISAAC images have limiting magnitudes of  24.8, 24.2 and 24.1 (10$\sigma$
for an extended source measured over 1\ffas0 diameter circular aperture) respectively.
These data were taken in 0\ffas4 seeing condition. Details about the optical
(BV{\it iz}) and near--IR observations and data reduction are given
in Giavalisco et al. (2004). The Spitzer--IRAC mid--IR images of the GOODS-S
are obtained in all 4 channels (3.6-8.0 $\mu$m) to 10$\sigma$ limiting
magnitudes for an isolated point source, from 25.8 (3.6 $\mu$m) to 23.0
(8.0 $\mu$m) magnitudes (Dickinson et al. in prep.).
Fluxes were measured in the MIPS data by fitting point sources to prior
positions of objects detected by IRAC, enabling reliable deblending even
in moderately crowded conditions (Chary et al., in prep.). The MIPS catalog
is 84\% complete at the formal 5$\sigma$ flux density limit (24~$\mu$Jy). 
In practice, detectability and photometric uncertainty in the IRAC and MIPS
data is ultimately a function of image crowding. We will visit this issue when
discussing the reliability of our candidates.

We block-averaged the ACS images (0\ffas03) to the same scale as that of the ISAAC
data (0\ffas15) and convolved them with a Gaussian approximation of the ISAAC PSF.
We then generated a source catalog by running SExtractor (Bertin \& Arnout 1996)
in dual image mode, using the ISAAC K--band as the detection image. A K--band selected
catalog was then constructed from the PSF--matched ACS (BV{\it iz}) and ISAAC
($JHK_s$) images, with total magnitudes (corresponding to MAG\_AUTO values from 
SExtractor) measured. 
Since the accuracy of near--IR photometry is crucial in
selecting and exploring the nature of the BBG candidates, we further examine
these by performing manual photometry on the ISAAC images of the BBG candidates.
The results from the two methods of photometry agree within their respective photometric
errors (see also Sect.~\ref{phot_error}). We coordinate--matched the K--band detected
sources with the weighted sum of channel 1 and 2 from the Spitzer--IRAC catalogs.
using SExtractor and PSFs appropriate for that channel (Dickinson et al. in prep.)
A maximum radial tolerance of 1$^{\prime\prime}$ was used to match sources
between the $K$--band and IRAC catalogs. We have found that matches with
larger separations are almost inevitably due to blending of multiple objects
in the IRAC images, which perturb the centroid position of the source as well
as corrupt the photometry, and are therefore to be avoided.

IRAC photometry was performed by measuring the magnitudes over
3$^{\prime\prime}$ or 4$^{\prime\prime}$ circular aperture diameters.
These were then converted to total magnitudes using aperture corrections
based on Monte Carlo simulations, in which artificial images of compact
galaxies (half--light radii $ < $0\ffas5) were added to the IRAC images
after convolving by the appropriate PSFs and subsequently recovered by
SExtractor.
The reason for using two different apertures is the potential for source
blending in the IRAC images. Blending may artificially brighten the IRAC
magnitude and hence force these sources into the selection range (Sect.~\ref{color}).
We used a 3$^{\prime\prime}$ aperture when estimating the IRAC photometry
for sources which have a nearest neighbor, measured in the K--band, within
a radius  $\leq$3\ffas0. If the separation was less than 1\ffas5 the source
was discarded. For the remaining sources we used a 4$^{\prime\prime}$ aperture.
The corrections to `total' magnitude are obtained from the simulations,
as described above, and are larger for the smaller aperture.
After selection of the final sample (Sect.~\ref{candidateparameters}), we
repeated the IRAC photometry through PSF fitting using the {\tt GALFIT}
package (Peng et al. 2002), and used these results in the SED fits.
This is discussed further in Sect~\ref{phot_error}.

\begin{figure}
\epsscale{0.9}
 \plotone{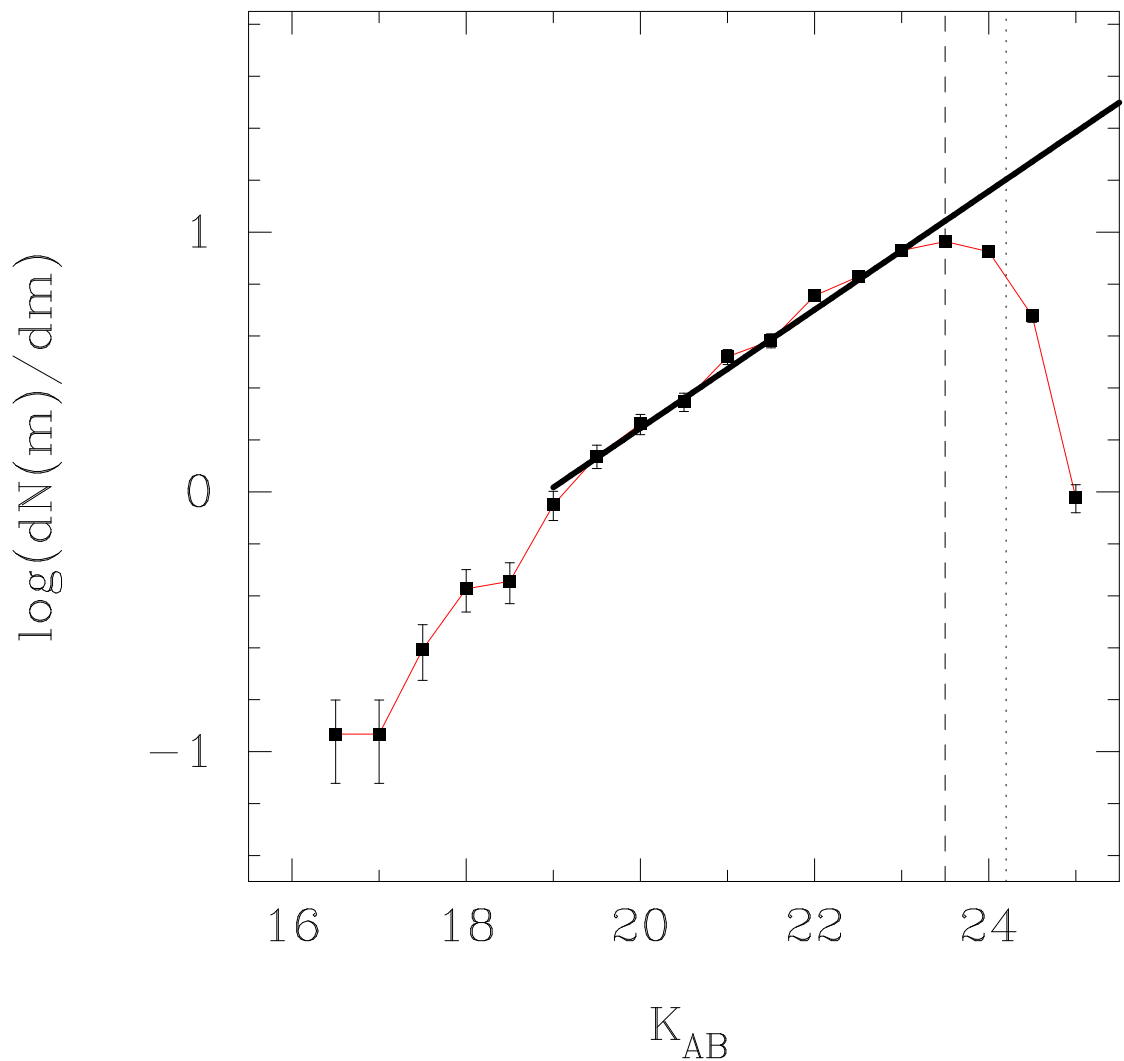}
\caption{Differential number counts of the $K_s$ magnitudes for sources in our
K-selected sample. The dashed vertical line represents the completeness limit of
$K_{\mathrm{AB}} \approx 23.5$ and the dotted vertical line represents the average $K_s$
magnitude of the BBG candidates.
A power-law function is fitted to the data for $19.0 < K_s < 23.5$ and is shown as a full-drawn black line.
The expected number of galaxies at $K_s = 24.21$ and the observed number are marked by red circles.
Their ratio is an estimate of the completeness at $K_s = 24.2$ (Sect.~\ref{completeness}).
}
\label{maglimits}
\end{figure}

The final result is a  K--band selected catalog containing total magnitudes in
ACS (BV{\it iz}), ISAAC ($JHK_s$) and IRAC ($m_{3.6}$, $m_{4.5}$, $m_{5.7}$,
$m_{8.0}$) bands. We estimate the completeness by fitting a power-law function
to the faint end of the differential number counts of the apparent K--magnitudes
for all galaxies in the K--selected catalog (Fig.~\ref{maglimits}). The catalog
is 82\% complete at $K_{AB}=23.5$. This catalog is used to identify candidates
satisfying our selection criteria.

\section{Selection of High Redshift Candidates}\label{selecting}

The selection and identification of evolved galaxies at $z \ga 5$ comprise two steps:
First, selection of likely candidates based on colors, and secondly, identification
of the most likely old and high redshift galaxies from these candidates by fitting SEDs
from population synthesis models.
This two-step process is necessary because, as we will show below, the colors
of post-starburst galaxies at $z \ga 5$ are to some extent degenerate with dusty
star forming galaxies at the same or lower redshifts.

\subsection{The Balmer break}\label{balmerbreak}

One feature in the spectral energy distribution (SED) of galaxies that can be used to identify
post-starburst galaxies at both high and low redshift is the Balmer break.
The Balmer break at 3648{\AA} is an age-dependent diagnostic of the stellar population.
The break is most prominent in A-stars (in O- and B-stars, the hydrogen is mostly ionized,
while in cooler late type stars, the opacity is dominated by H$^-$, with a maximum
opacity at 8500{\AA}).
For a single generation of stars, the break is most pronounced for ages between 0.1$-$1.0
Gyr. However, the development of the Balmer break occurs for stellar populations in both
passively evolving and continuous star formation scenarios, but on different time scales.
For an instantaneous star burst, followed by passive evolution, the break develops
when O- and B-stars leave the main sequence, and for continuous star formation, when the
number of O-stars have reached a more or less constant value while the number of A-stars
is still increasing (cf. Leitherer et al. 1999).
The Balmer break has the potential to resolve the age--extinction degeneracy.
Most extinction laws have a relatively smooth dependence on wavelength
and will not produce the step-like feature of the Balmer break. Its usefulness,
however, is limited by the photometric accuracy relative to the amplitude of the
3648{\AA} break.

In this paper we concentrate on galaxies at $z \ga 5$. For redshifts in the range
$z \approx 5 - 9$, the Balmer break is located between the K- and 3.6$\mu$m
passbands. In this redshift range, observed optical wavelengths correspond to the
extreme UV region, which is mostly lost, through the Lyman--break, interstellar
and intergalactic absorption.
This means that the selection of $z \ga 5$ galaxies is greatly aided by using observed
near- and mid-infrared wavelengths. Such selection criteria have only become possible
with the availability of relatively deep imaging with the IRAC instrument on Spitzer.

\subsection{Stellar population synthesis models}\label{models}

Stellar population synthesis models will be used for two purposes.
First, the models are used to define regions in color-color plots
which are the likely location of $z \ga 5$ post--starburst candidates.
This is done by defining a limited set of parameters characterizing
this type of galaxy and following their color evolution as a function
of redshift.
Secondly, the models are used to fit  the observed broad--band photometric
data of the color-selected candidates. Apart from providing global galaxy parameters,
such as redshift, age and stellar mass, this will allow a clear distinction between the
type of galaxies in which we are interested and  interlopers of various kinds.

We use the stellar population synthesis models of Bruzual and Charlot (2003; BC03)
to explore the broad--band color evolution of galaxies with different star formation
histories, ages and metallicities.
%%%
In order to fit the SED of each galaxy in an unbiased and prior-free manner, we
explore a large parameter space for redshift, stellar age, extinction, metallicity
and star formation history. While hidden priors cannot be avoided due to the cut--offs
in parameter values, as well as the form assumed for the star formation history,
we strive to keep these to a minmum. The number of parameters used to define the
SED is ultimately limited by the number of photometric data points.

We use a Salpeter initial mass function,
(IMF) with lower and upper mass cut-offs at 0.1 and 100 M$_{\odot}$, respectively.
The resulting spectral energy distributions are redshifted in the range $z=0.2-8.6$
with $\Delta z = 0.1$, and their colors are evaluated in fixed observed bands (ACS:
BV{\it iz}; ISAAC/VLT: JHK$_{\mathrm{s}}$; IRAC/Spitzer: 3.6, 4.5, 5.7, 8.0$\mu$m).
We do not include longer wavelength MIPS data in the fitting process as the
BC03 models do not include dust or PAH emission.
Dust obscuration is parametrized using the attenuation law of Calzetti et al. (2000).
It is parametrized through the E$_{\mathrm{B-V}}$ color index, covering the range
E$_{\mathrm{B-V}} = 0.0-0.95$, with $\Delta\mathrm{E}_{\mathrm{B-V}} = 0.025$.
Additional attenuation is introduced through neutral hydrogen absorption in the intergalactic
medium (IGM). We used the Madau (1995) prescription for the mean IGM opacity.
The age of a stellar population is measured from the onset of star formation.
We adopt simple monotonic star formation histories, as with the present photometric
data we cannot quantitatively assess the goodness of fit for models with multiple
previous bursts of star formation. Although this may influence the estimates of the
average stellar age, obscuration and total stellar mass\footnote{As shown in Papovich
et al. (2006), models which incorporate multiple bursts of star formation may result
in larger derived stellar masses.}, it does not substantially affect the overall shape of
the SED. Since the photometric redshift is based on distinct features in the SED, it is
a robust estimate regardless of the stellar populations considered.
The age range extends from 5 Myr to 2.4 Gyr, with steps of 5 Myr
up to 100 Myr, followed by age steps of 100 Myr up to 2.4 Gyr. The maximum age
corresponds to the age of the universe at $z \approx 2.7$.
Four different metallicities are used, 0.2, 0.4, 1.0, 2.5 Z$_{\odot}$.
The star formation history is parametrized as an exponentially decreasing
star formation rate, where $\tau$ represents the e--folding decay time. We use
$\tau = 0.0, 0.1, 0.2, 0.3, 0.4, 0.6, 0.8$ and $1.0$ Gyr. The $\tau = 0$
case represents an instantaneous starburst.

A large number of models ($\sim 2.5 \times 10^6$) are pre--computed, spanning
the parameter space as defined above. The resulting spectral energy distributions
are integrated through the appropriate filter response functions, We also derive
a bolometric luminosity by integrating over the entire wavelength range.
Finding the best--fit parameters for a given set of photometric data points then
involves normalization to the observed fluxes and calculation of the goodness-of-fit
for each point in parameter space. The best--fit model parameters are selected from
the model resulting in the minimum $\chi^2$. Since a $\chi^2$ value is derived
for all parameter combinations, the confidence of the fit can easily be evaluated.
The SED fitting is done using flux densities ($f_{\nu}$). The treatment of observed
upper limits needs special consideration. The model SEDs have extremely steep
flux density gradients at wavelengths shorter than 1216\AA, and have essentially
zero flux below the Lyman limit at 912\AA. If the redshift is high enough to shift the
Lyman limit to wavelengths redder than a given filter, the observed upper limit becomes
useless as the difference between the upper limit flux density and that given by the
model can amount to several orders of magnitude. On the other hand, if the redshift
is low enough to place the filters with upper flux limits on the red side of the Lyman
limit, the upper limit has a more meaningful role in constraining the model fit.
Due to the large difference between the observed flux limits and the model flux for
high redshift objects, where the Lyman limit is on the red side of the upper limit,
the $\chi^2$ estimate will invariably favor a lower--z solution, but with a very poor
fit both at short and long wavelengths, and a correspondingly large $\chi^2$ value.
This introduces a bias, which we overcome by not including the upper limits in the
$\chi^2$ estimate whenever the model SED is fainter than the upper limit. However,
when the flux of the model SED is larger than the observed upper limit, thus violating
an observed constraint, we include this in the $\chi^2$ estimate.

\subsection{Monte Carlo simulations}\label{montecarlo}

%%%
With a limited set of photometric data points, it is necessary to keep the number
of model parameters to a minimum in order to achieve a meaningful goodness--of--fit
estimate. In addition, degeneracies between some of the parameters, such as stellar age,
extinction and metallicity exist and can potentially lead to a large area of parameter
space where a good fit between the model and the observed data points can be found.
One way out of this dilemma is to apply priors, where we assume certain properties
of the galaxies being fitted. While this can lead to a `sharper' solution, it also
carries the potential of introducing biases. We have chosen to keep the
priors to a minimum (Sect.~\ref{models}) and accept a somewhat more diffuse solution
space for a few cases but keeping the solutions as unbiased as possible.

In order to define the confidence and test the stability of the model fitting and
the resulting solution space, we performed Monte Carlo simulations, where the fluxes
in all photometric bands are allowed to vary simultaneously within their nominal errors.
The errors are assumed to normally distributed and uncorrelated. While these assumptions
are only partly true, due to sensitivity limits and zero--point uncertainties in the
photometry, they represent a good approximation to the true photometric uncertainty.
The resulting distribution of the best--fit values for each parameter represent the
probability distribution for this particular parameter. As we will see, a small
percentage of the Monte Carlo realizations result in a best--fit at a lower redshift.
The actual fit of these solutions can be good, but represents an unlikely combination
of the observed photometric data values, given their errors.

%An additional, and in many ways superior, test of the confidence and stability of the
%model fitting can be obtained through Monte Carlo simulations, where the fluxes in
%all bands are allowed to vary simultaneously within their nominal errors. The errors
%are assumed to be normally distributed and uncorrelated. While these assumptions are
%only partly true, due to sensitivity limits
%and zero--point uncertainties in the photometry, they represent a good approximation to
%the true photometric uncertainty.
%
%For our models,
We generate $10^3$ realizations of the photometric data set for each galaxy. In each
realization we allow each photometric data point to vary stochastically as described above.
The bands with non--detections are still treated as upper limits. We then determine the
best--fit parameters for each realization of the photometric data in the same manner as
described in Sect.~\ref{models}. The resulting distribution of redshift, age, stellar mass,
extinction, etc. for the 10$^3$ Monte Carlo realizations allows an estimate of the
confidence of the various solutions. This gives a more accurate estimate of the confidence
% represent probability distributions for the
%model parameters.
%This allows an estimate of the  confidence level with better quality
than a single realization and the corresponding variation of the $\chi^{2}$ values.

\subsection{The color selection technique}\label{color}

We are primarily interested in galaxies with a well defined Balmer break,
i.e. with ages $> 0.2$ Gyr, and situated at redshifts $z\ga5$. Hence, the primary
color parameter is the K$_{\mathrm{s}}-3.6\mu$m color, which straddles the
3648{\AA} Balmer break at $5 < z < 9$ (see Sect.~\ref{balmerbreak}).

\begin{figure}
\epsscale{1.0}
 \plotone{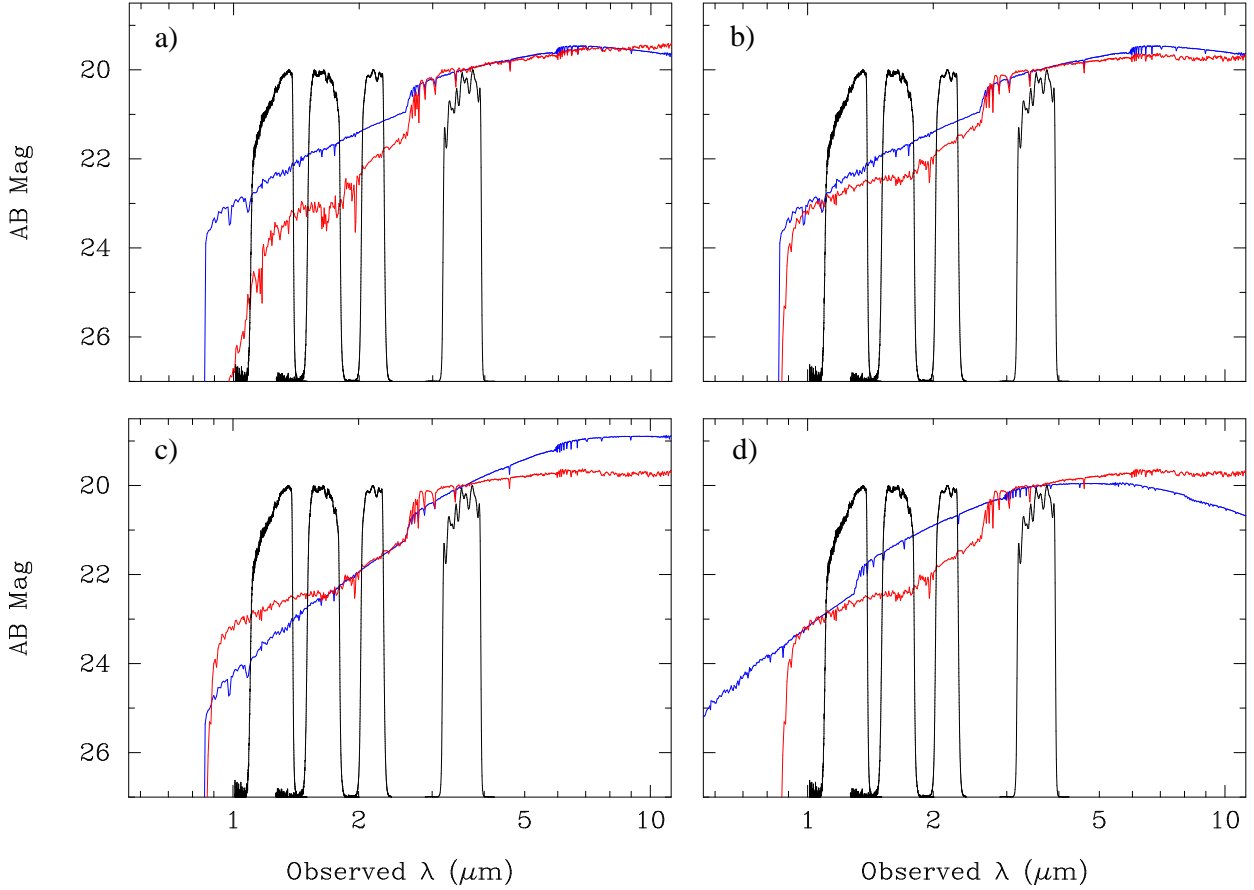}
\caption{Examples of SEDs derived from the Bruzual \& Charlot 2003 models,
illustrating the effects of various parameters and their impact on the photometry.
Each panel shows a post-starburst galaxy (red) and a dusty starburst (blue),
with the response curves for the ISAAC JHK$_{\mathrm{s}}$ and IRAC
3.6$\mu$m filters overlaid.
The SED of the post-starburst / dusty starburst galaxies have the
following properties:
{\it a)}\ $z = 6.0/6.0$, E$_{\mathrm{B-V}} = 0.0/ 0.5$, age $= 600/5$ Myr, $Z = 1.0/1.0$\,Z$_{\odot}$,
{\it b)}\ $z = 6.0/6.0$, E$_{\mathrm{B-V}} = 0.0/ 0.5$, age $= 600/5$ Myr, $Z = 0.2/1.0$\,Z$_{\odot}$,
{\it c)}\ $z = 6.0/6.0$, E$_{\mathrm{B-V}} = 0.0/ 0.7$, age $= 600/5$ Myr, $Z = 0.2/1.0$\,Z$_{\odot}$,
{\it d)}\ $z = 6.0/2.5$, E$_{\mathrm{B-V}} = 0.0/ 0.7$, age $= 600/5$ Myr, $Z = 0.2/1.0$\,Z$_{\odot}$.
Hence, the SED for the post-starburst model is the same in panels
b), c) and d), while the SED of the dusty starburst model is the same in
panels a) and b).}
\label{comp_sed}
\end{figure}

There are several physical parameters that can cause red colors in a stellar population,
including the age of the stellar population, metallicity and dust extinction. Because of
this degeneracy, a single color is usually not a robust indicator of redshift, nor
does it distinguish between different galaxy types; for instance, an obscured star
forming galaxy, a post--starburst galaxy and an elliptical galaxy.
This is illustrated in Fig.~\ref{comp_sed}, where we show the SED for a
typical post--starburst (red line) and a dusty starburst galaxy (blue line)
for a variety of age, metallicity and E$_{\mathrm{B-V}}$ parameters.
The model galaxies are placed at $z = 6.0$, except
in one case where the dusty starburst is located at $z = 2.5$. In  addition
to the SEDs, we also show the ISAAC JHK$_{\mathrm{s}}$ and the IRAC
3.6$\mu$m bandpasses. All the SEDs are normalized at 3.6$\mu$m.
In Fig~\ref{comp_sed}a, both the post-starburst and dusty starburst are at
$z = 6.0$, both have solar metallicity, with the only difference being
their age (600 Myr vs. 5 Myr) and the extinction E$_{\mathrm{B-V}}$ (0.0
vs 0.5). In this case, the SED of the post-starburst galaxy has a larger
gradient at wavelengths shorter than the Balmer break compared to the dusty
starburst. Therefore, in this case it is possible to distinguish between
these two galaxy types by using a second color index.
However, in Figs.~\ref{comp_sed}b and c, we demonstrate the effect
when relatively small changes to the galaxy parameters are incorporated.
In Fig~\ref{comp_sed}b the metallicity of the post-starburst galaxy
is decreased to 0.2Z$_{\odot}$, resulting in a somewhat less steep SED gradient
short-ward of the Balmer break. In Fig.~\ref{comp_sed}c, in addition to
the lower metallicity of the post-starburst galaxy, the E$_{\mathrm{B-V}}$
of the dusty starburst increased to 0.7. In this case, the dusty starburst
galaxy has a steeper gradient shortwards of the Balmer break than the
post-starburst. Finally, in Fig.~\ref{comp_sed}d, we keep the parameters
the same as in Fig.~\ref{comp_sed}c but move the dusty starburst galaxy
to $z = 2.5$.
The SEDs in Fig.~\ref{comp_sed} show that, in general, even the use of
two color indices may not be sufficient to distinguish between post-starburst
and dusty starburst galaxies. It is, however, possible to identify and remove
elliptical galaxies from the sample.
While the K$_s-3.6\mu$m color index is the main parameter used for
selecting post--starburst galaxies at $z \ga 5$, the number of interlopers
can be minimized by using a second color index.
In this paper  we explore the use of the J$-$K$_s$, as well as the H$-3.6\mu$m
colors as a secondary index.

\begin{figure}
\epsscale{1.0}
 \plotone{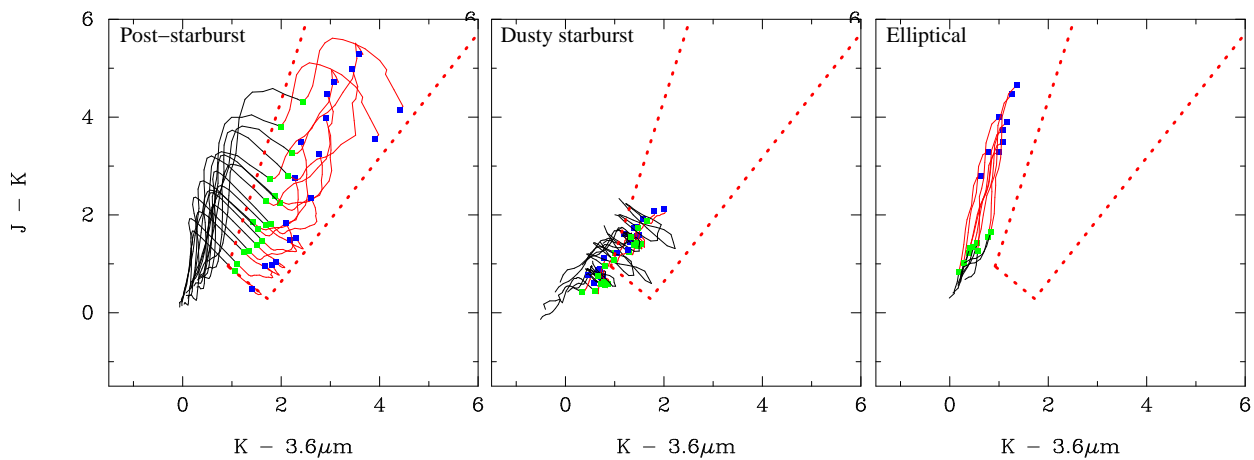}
\caption{Tracks of three different types of model galaxies in the J$-$K and
K$_s - 3.6\mu$m color plane. The model tracks represent a range of parameters
in E$_{\mathrm{B-V}}$, age, and star formation history, characteristic for
post-starburst, dusty starburst and elliptical galaxies. The ranges of the parameters
are given in Table~\ref{parameters}. Each track starts at $z = 1$ and extends
to $z = 8$, with green dots representing $z = 5$ and blue dots $z = 8$, except for
the elliptical galaxies where the corresponding redshifts are $z = 2$ and
$z = 4$. The region inside the wedge outlined by the dashed red line correspond
to region A, where we expect to find $z > 5$ post-starburst galaxies (BBGs).}
\label{tracks1}
\end{figure}

\begin{figure}
\epsscale{1.0}
 \plotone{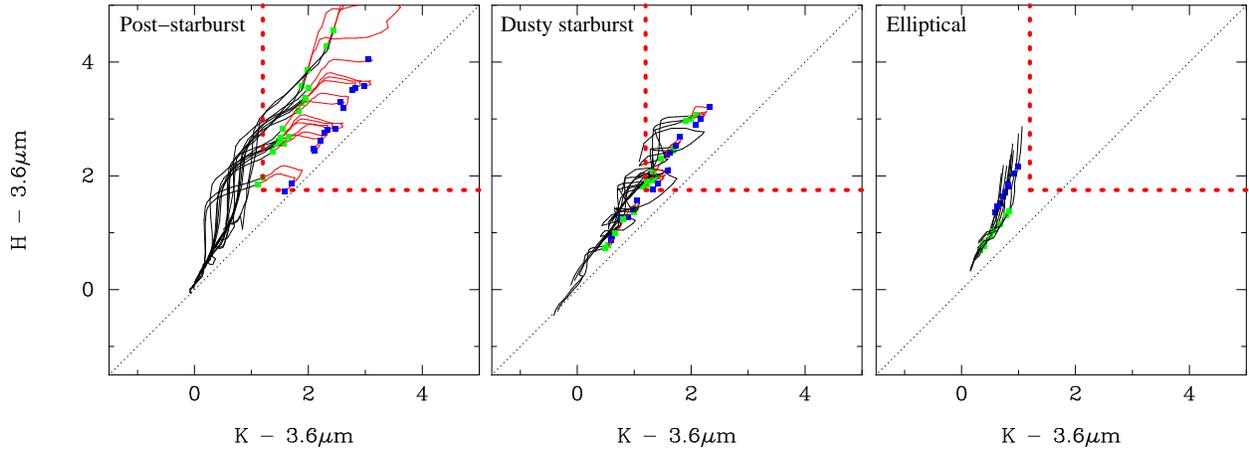}
\caption{Same as Fig.~\ref{tracks1} for H$-3.6\mu$m vs K$_s - 3.6\mu$m. The region
above and to the right of the dashed red line defines the region of post--starburst
candidates at $z>5$. These color indices define region B (see text).}
\label{tracks2}
\end{figure}

In order to better understand the behavior of different types of galaxy
models when using two color indices, and to explore their limitations,
we constructed synthetic galaxy SEDs for a set of post-starburst,
dusty starburst and elliptical galaxies using the Bruzual \& Charlot
(2003) models. The models explore a wide range of parameter
combinations appropriate for each galaxy type (see Table~\ref{parameters}).
Broad-band photometric data were obtained by convolving the SEDs
with the appropriate filter response functions.
In Figures~\ref{tracks1} and~\ref{tracks2} we show the resulting tracks
when each galaxy model, for a fixed set of parameters, is shifted
to different redshifts. For the post--starburst and dusty starburst
models, the redshift ranges from $z=1-8$.
Tracks at $z < 1$ do not overlap the ones at higher redshift and have
been omitted from the figures.
Each track is marked with
a green and blue dot, corresponding to $z = 5$ and $z = 8$, respectively.
The elliptical models are restricted to the range $z=1-4$, with the green and blue dots
marking $z = 2$ and $z = 4$.

Using the post-starburst tracks, and limiting the redshift to $5 < z < 8$,
we can define a region on the color-color plane which contains all
of the model tracks. This is done for both the J$-$K vs. K$_s - 3.6\mu$m
and H$- 3.6\mu$m vs. K$_s - 3.6\mu$m indices (Figs.~\ref{tracks1} -- region A
and~\ref{tracks2} -- region B, respectively). While the tracks for elliptical
galaxies fall well outside the regions defining the post--starburst tracks,
this is not the case for dusty starburst galaxies which occupy a region
overlapping with the post--starburst galaxies. The best way of separating
these types is to introduce more constraints by fitting the SEDs over the
entire wavelength range available and select post--starburst galaxies based
on their respective model parameters.

The expected location of post--starburst galaxies on the J$-$K vs.
K$_s-3.6\mu$m plane is defined by (see Fig.~\ref{tracks1} -- region A):
\begin{eqnarray}
\mathrm{J}-\mathrm{K} & < &  -1.94 + 3.14\,(\mathrm{K}_s-3.6\mu\mathrm{m})\ \ \mathrm{\&} \nonumber \\ 
\mathrm{J}-\mathrm{K} & > &  -1.90 + 1.27 (\mathrm{K}_s - 3.6\mu\mathrm{m})\ \ \, \mathrm{\&} \nonumber \\
\mathrm{J}-\mathrm{K} & > &   1.71 - 0.82\,(\mathrm{K}_s - 3.6\mu\mathrm{m}) \nonumber
\end{eqnarray} 
For the case of H$-3.6\mu$m vs. K$_s-3.6\mu$m, the region of interest is defined as
(see Fig.~\ref{tracks2} -- region B):
\begin{eqnarray}
\mathrm{H}-3.6\mu\mathrm{m}  >  1.75
& \mathrm{\&} &
\mathrm{K}_s-3.6\mu\mathrm{m} > 1.20 \nonumber
\end{eqnarray}

\section{Results}\label{results}

\subsection{The Balmer--break candidates}

\begin{figure}
\epsscale{1.0}
 \plotone{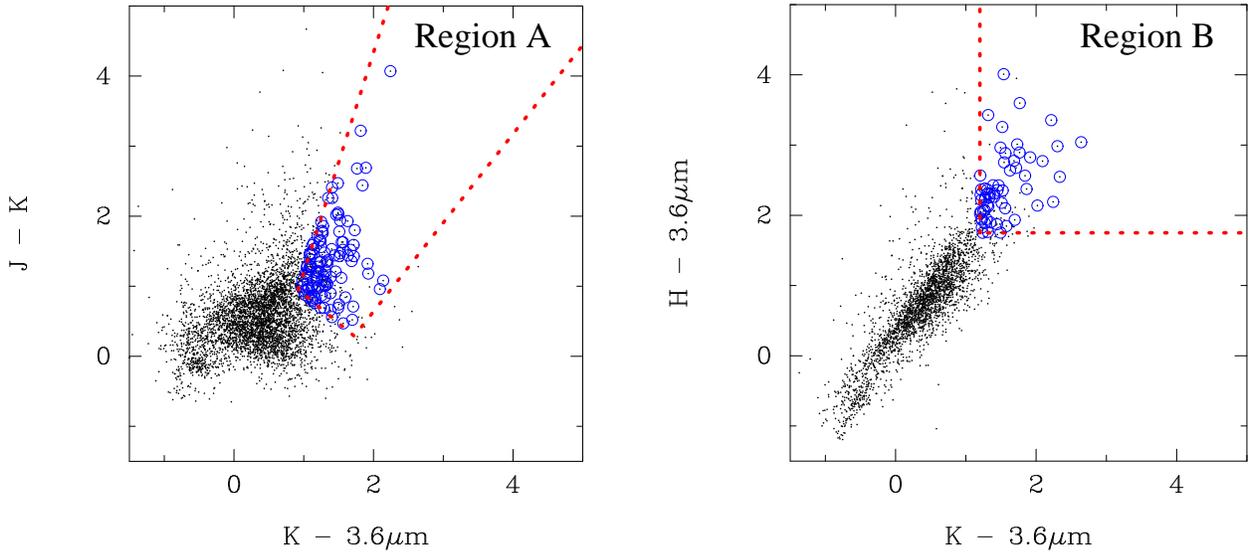}
\caption{The color indices for the K$_s$-selected sample from the GOODS sample.
Two alternative methods for selecting high redshift post-starburst galaxies
using near- and mid-infrared colors. The objects located within the selection region
but lacking a ring are detected in the B--band and were excluded from the color
selection (cf. Sect.~\ref{results}).
{\bf Region A:}\ J$-$K vs. K$_s - 3.6\mu$m: the area inside the wedge outlined by the red line
contains $z>5$ post-starburst and dusty starburst galaxies (see Fig~\ref{tracks1}).
The selected candidates (shown as circles) will also contain dusty starburst systems
at redshifts $z \approx 2-8$.
{\bf Region B:} H$-3.6\mu$m vs. K$-3.6\mu$m: the area above and to the right of the red line
contains $z\ga5$ post-starburst and dusty starburst galaxies (see Fig~\ref{tracks2}).
}
\label{colorcolor}
\end{figure}

We select $z \ga 5$ candidates from our K$_s$--selected catalog using the
color--color diagrams shown in Fig.~\ref{colorcolor}. 
In order to limit the number of selected sources, we also required them to
be undetected in the $B$--band, with $m(B_{435} > 27.85$. At $z > 4.3$,
the 912\AA\ Lyman limit redshifts entirely redward of the ACS F435W filter
bandpass, and this requirement lowers the number of foreground objects
included in the selection\footnote{In a few cases the catalog value for the
B--band would be less than 27.8, but with an uncertainty $>$1.0 mag. If this
was the case, we regarded it as an upper limit.}. 

We call the selection based on the $J-K$ vs. $K_{\mathrm{s}} - 3.6\mu$m
colors for region {\it A}, and that based on $H - 3.6\mu$m vs.
$K_{\mathrm{s}} - 3.6\mu$m for region {\it B}.
As noted in Sect.~\ref{data}, we also require that the ISAAC and IRAC centroid
positions do not differ by more than 1\ffas0. We find that larger offsets
inevitably indicate problems with blending in the IRAC data. On the other
hand, there are small, residual astrometric distortions in the current public
data products for both the ISAAC and IRAC GOODS images at levels of up to
0\ffas4 which make it impractical to adopt much smaller matching tolerances.
In the final selection, the center positions of the ISAAC and IRAC sources
generally agree to better than 0\ffas5, although there are two exceptions
(BBG\#3361, 0\ffas6, and BBG\#2068, 1\ffas0; see Table~\ref{table2}).

Regions A and B contain 112 and 60 Êsources, respectively. ÊAs noted in
Sect.~\ref{data}, the ISAAC $H$-band images cover a Êsmaller solid angle
(124 arcmin$^2$) than those at $J$ and $K_s$ (159 arcmin $^2$), which
partially accounts for the smaller number of sources in color Êregion B.
There are 38 objects that are common to the two color selection regions.
The previously identified J--band drop--out galaxy, JD2, found in the HUDF
by Mobasher et al. (2005), is contained in both region A and B (BBG\#3179).

Fitting Bruzual \& Charlot models to all sources in regions
{\it A} and {\it B} shows that 9 from region {\it A} and 8 from region
{\it B} have photometric redshifts $z \ga 5$ and ages $>0.2$ Gyr.
Four of the high redshift candidates are common to both regions.
The remaining sources in regions {\it A} and {\it B} have best--fit solutions
consistent with dust--obscured starburst galaxies at redshifts $z \approx 1-3$
or dusty post--starburst galaxies at $z < 4 $.

Combining sources in region {\it A} and {\it B}, which include the previously
detected source HUDF--JD2 (BBG\#3179), we have a sample of 13 high
redshift Balmer--break candidates. The coordinates and photometric data of the
13 candidates found here are given in Table~\ref{table1} (the data for BBG\#3179/JD2
are taken from Mobasher et al. 2005, but see caption to Table~\ref{table1} for a
revision of the photometry).

\begin{figure}
\epsscale{0.85}
 \plotone{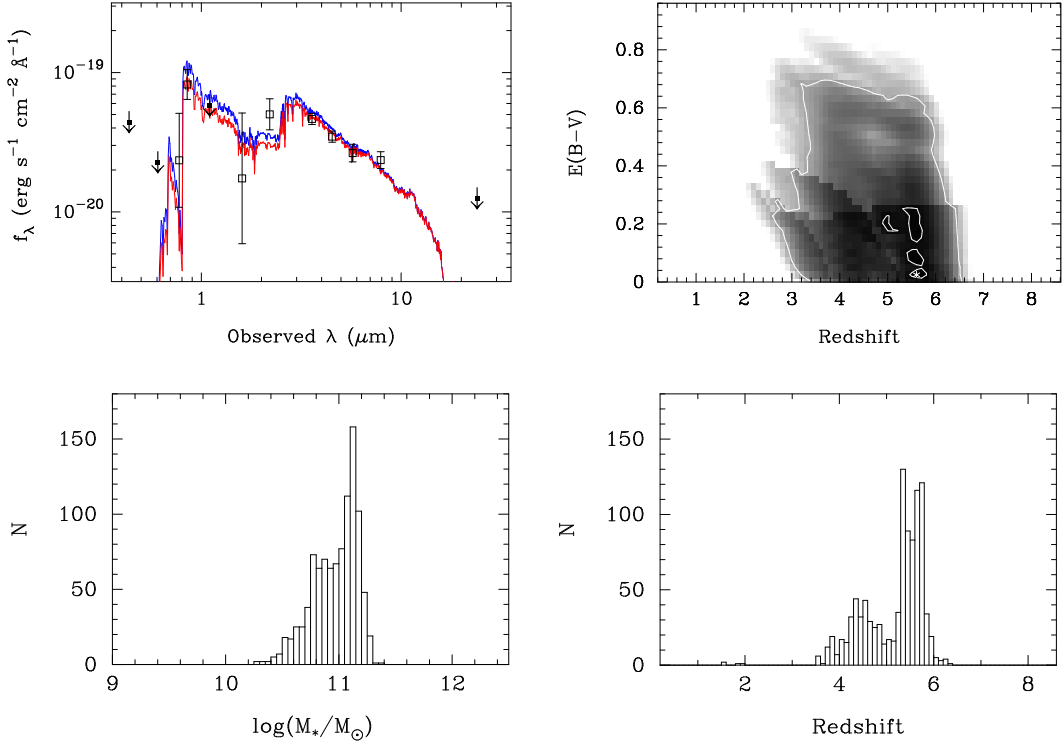}
\caption{
{\it Top:}\ Images of BBG\#547. Starting from the top left, the panels
show the ACS BV{\it iz} bands, the ISAAC/VLT JHK$_s$ bands, the Spitzer
IRAC 3.6, 4.5, 5.8, 8.0$\mu$m bands and, finally, the Spitzer MIPS 24$\mu$m image.
The middle left panel ({\it a)}, shows the observed data with the best-fit model SED. The red line
shows the SED with extinction (when present). The blue line (when present) shows the SED
corrected for dust extinction.  The MIPS 24$\mu$m data is shown but is not used in the
fitting procedure.
The middle right panel ({\it b)}, shows contours of $\chi^{2}_{\nu}$
values for the best fit as a function of redshift and extinction E$_{\mathrm{B-V}}$.
Bottom panels show the results of 10$^3$ Monte Carlo realizations for redshift and
stellar mass.
These panels represent the probability distribution of the two parameters.
The best-fit parameters are given in
Table~\ref{table2}:
$z = 5.6$; E$_{\mathrm{B-V}} = 0.025$; age $=$ 0.8 Gyr; $\tau$ = 0.2 Gyr; Z $=$ 2.5\,Z$_{\odot}$;
log(M$_{*}$/M$_{\odot}) = $11.056.}
\label{all_0547}
\end{figure}

\begin{figure}
\epsscale{0.85}
 \plotone{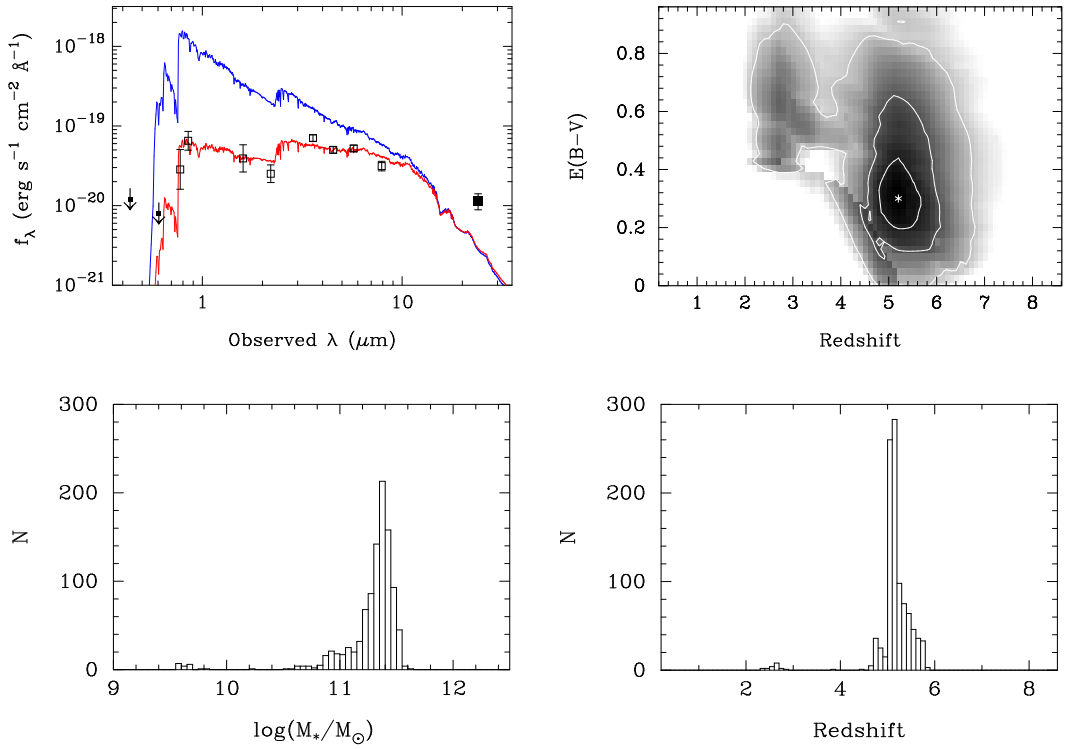}
\caption{Same as Fig.~\ref{all_0547} for BBG\#2068:
$z = 5.2$; E$_{\mathrm{B-V}} = 0.300$; age $=$ 1.0 Gyr; $\tau$ = 0.8 Gyr; Z $=$ 2.5\,Z$_{\odot}$;
log(M$_{*}$/M$_{\odot}) = $11.298.}
\label{all_2068}
\end{figure}

\begin{figure}
\epsscale{0.85}
 \plotone{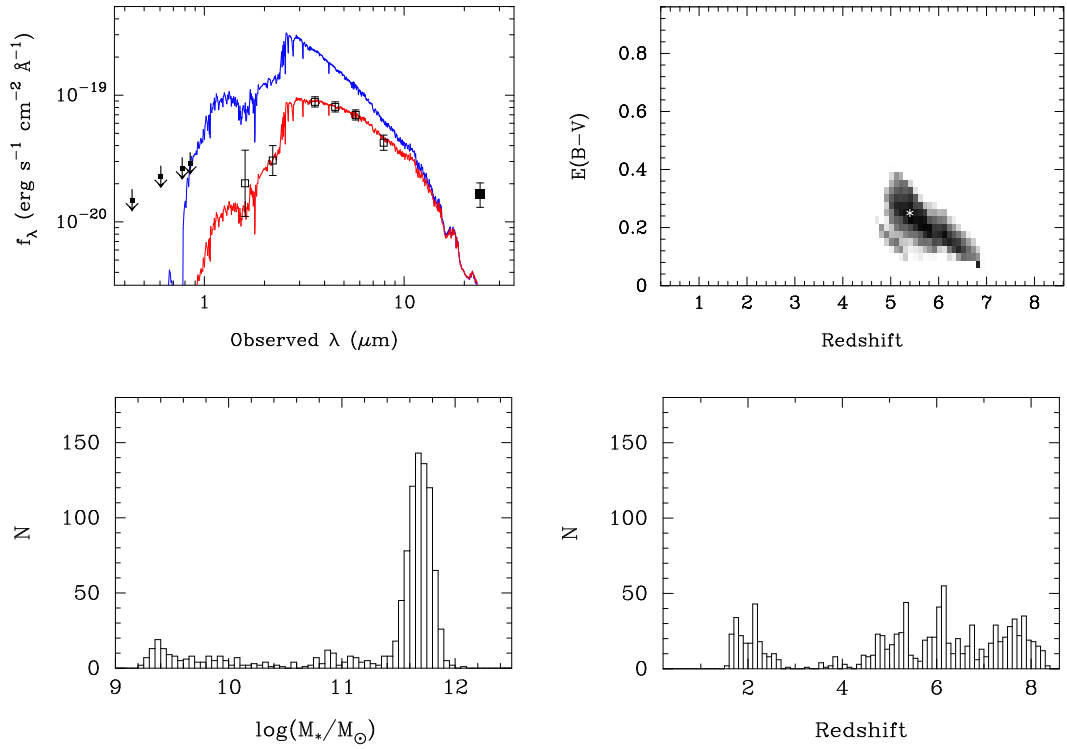}
\caption{Same as Fig.~\ref{all_0547} for BBG\#2864:
$z = 5.4$; E$_{\mathrm{B-V}} = 0.250$; age $=$ 0.9 Gyr; $\tau$ = 0.1 Gyr; Z $=$ 0.2\,Z$_{\odot}$;
log(M$_{*}$/M$_{\odot}) = $11.63.}
\label{all_2864}
\end{figure}

\begin{figure}
\epsscale{0.85}
 \plotone{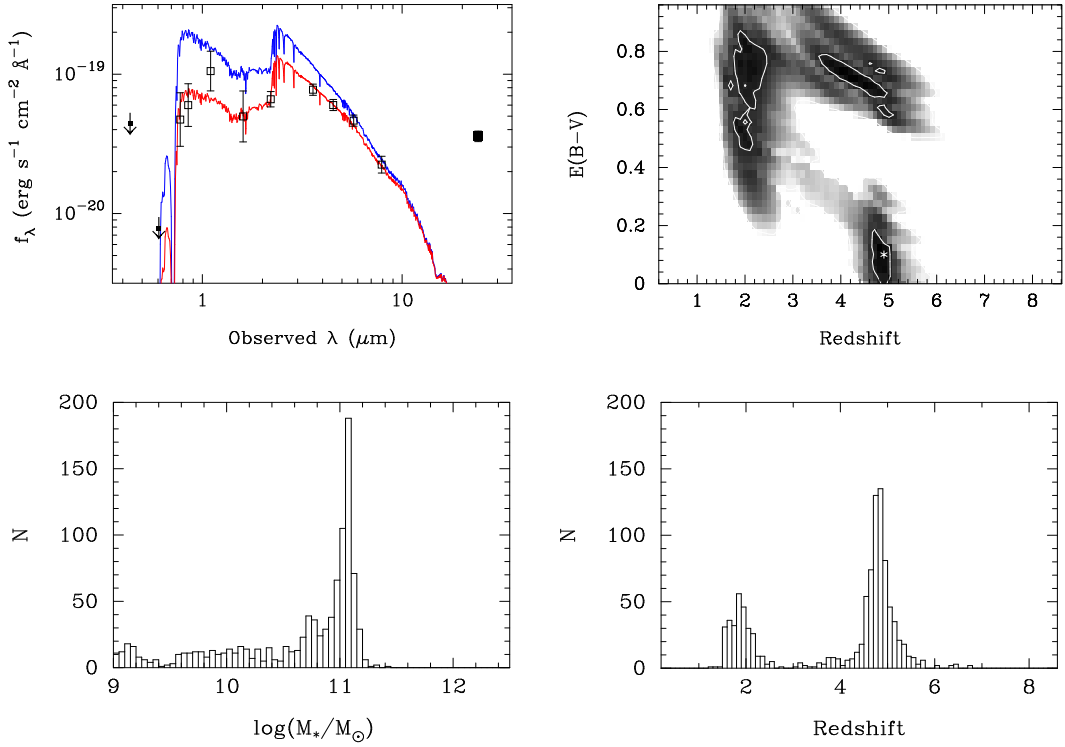}
\caption{Same as Fig.~\ref{all_0547} for BBG\#2910:
$z = 4.9$; E$_{\mathrm{B-V}} = 0.100$; age $=$ 0.4 Gyr; $\tau$ = 0.0 Gyr; Z $=$ 0.2\,Z$_{\odot}$;
log(M$_{*}$/M$_{\odot}) = $11.053.}
\label{all_2910}
\end{figure}

\begin{figure}
\epsscale{0.85}
 \plotone{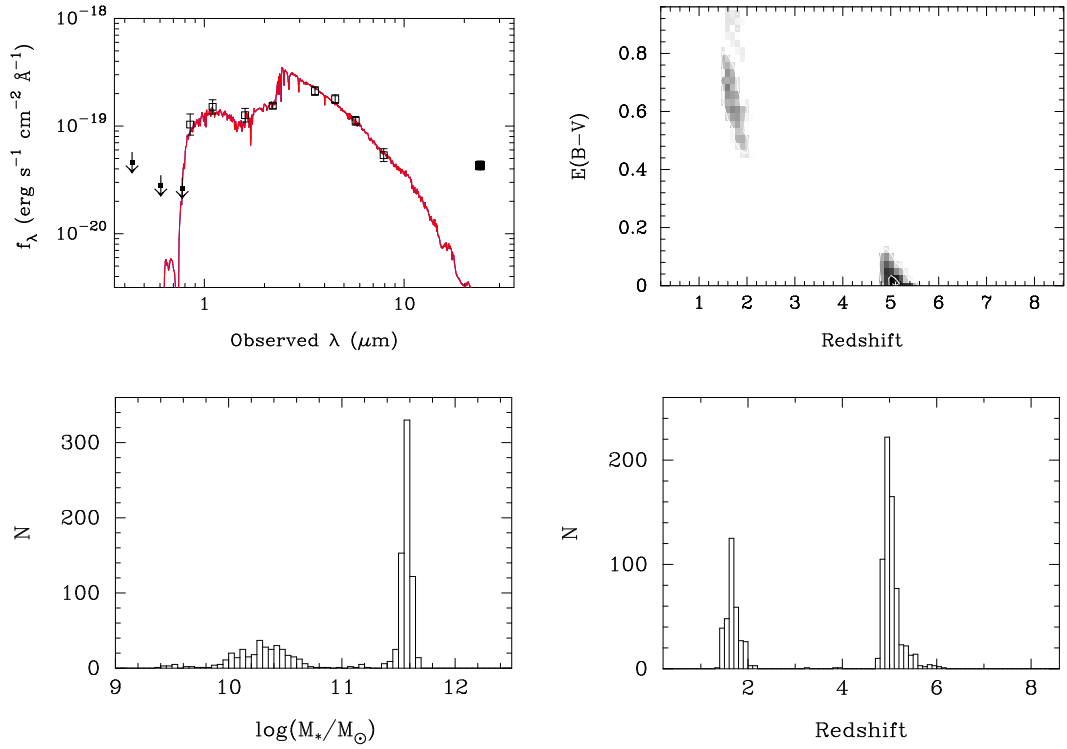}
\caption{Same as Fig.~\ref{all_0547} for BBG\#3348:
$z = 5.1$; E$_{\mathrm{B-V}} = 0.0$; age $=$ 0.9 Gyr; $\tau$ = 0.1 Gyr; Z $=$ 0.2\,Z$_{\odot}$;
log(M$_{*}$/M$_{\odot}) = $11.573.}
\label{all_3348}
\end{figure}

\begin{figure}
\epsscale{0.85}
 \plotone{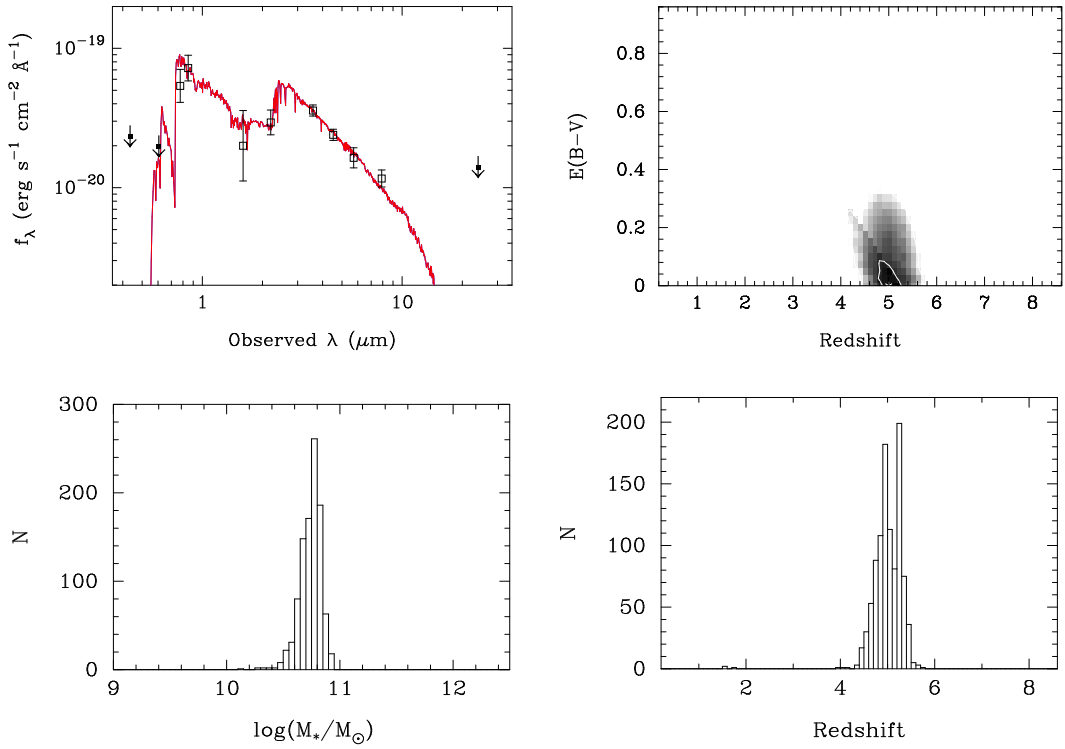}
\caption{Same as Fig.~\ref{all_0547} for BBG\#3361:
$z = 5.0$; E$_{\mathrm{B-V}} = 0.000$; age $=$ 0.8 Gyr; $\tau$ = 0.2 Gyr; Z $=$ 1.0\,Z$_{\odot}$;
log(M$_{*}$/M$_{\odot}) = $10.751.}
\label{all_3361}
\end{figure}

\begin{figure}
\epsscale{0.85}
 \plotone{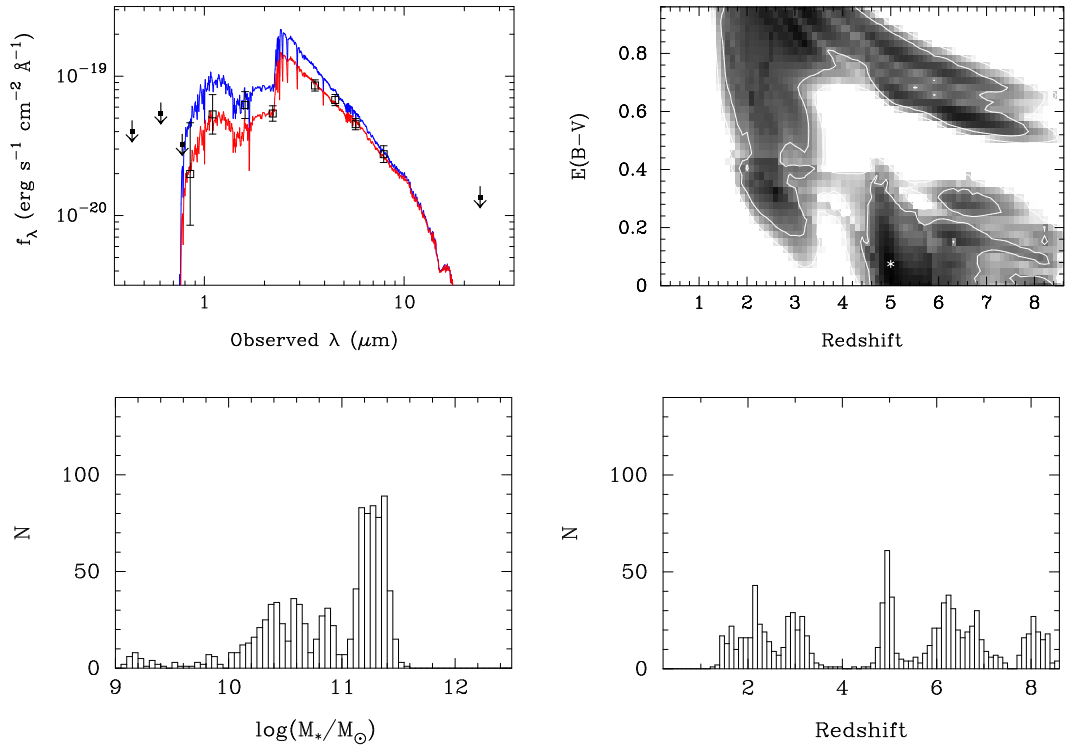}
\caption{Same as Fig.~\ref{all_0547} for BBG\#4071:
$z = 5.0$; E$_{\mathrm{B-V}} = 0.075$; age $=$ 0.4 Gyr; $\tau$ = 0.0 Gyr; Z $=$ 1.0\,Z$_{\odot}$;
log(M$_{*}$/M$_{\odot}) = $11.172.}
\label{all_4071}
\end{figure}

\begin{figure}
\epsscale{0.85}
 \plotone{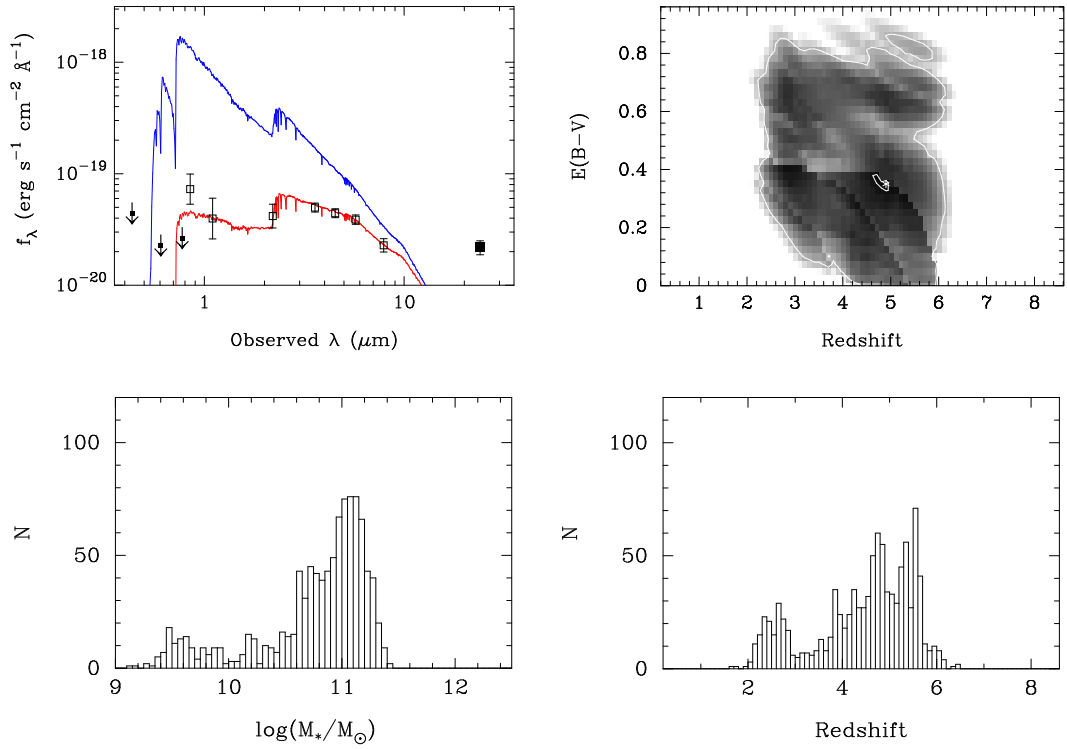}
\caption{Same as Fig.~\ref{all_0547} for BBG\#4135:
$z = 4.9$; E$_{\mathrm{B-V}} = 0.350$; age $=$ 0.3 Gyr; $\tau$ = 0.1 Gyr; Z $=$ 0.2\,Z$_{\odot}$;
log(M$_{*}$/M$_{\odot}) = $10.928.}
\label{all_4135}
\end{figure}

\begin{figure}
\epsscale{0.85}
 \plotone{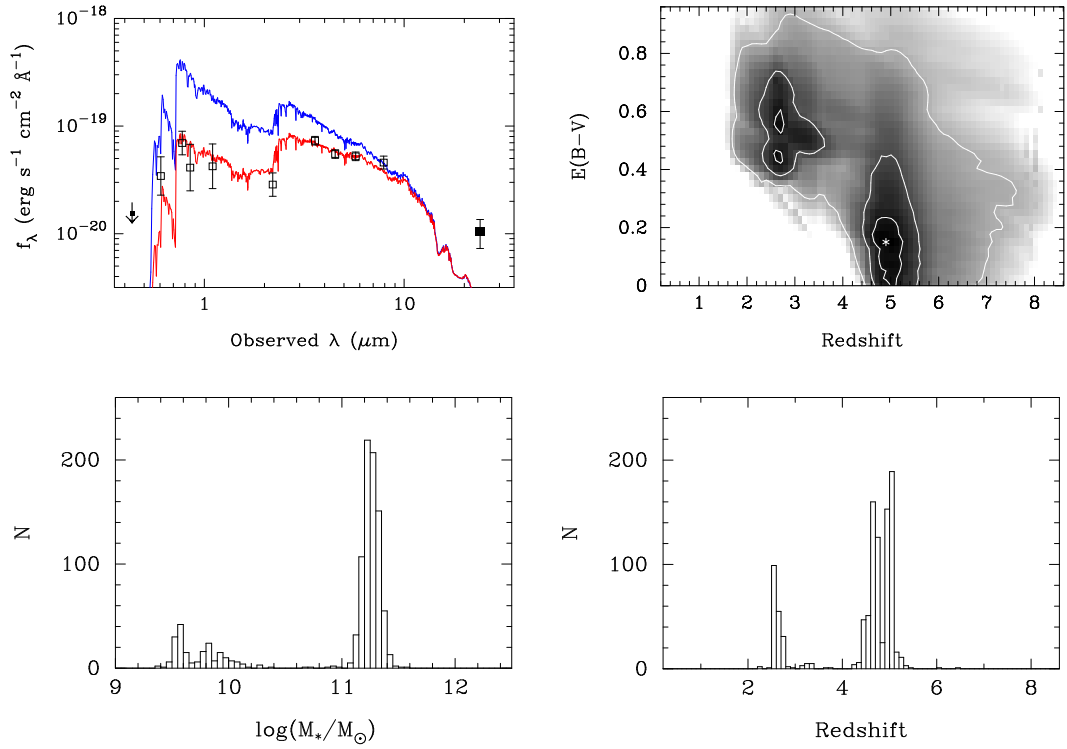}
\caption{Same as Fig.~\ref{all_0547} for BBG\#4550:
$z = 4.9$; E$_{\mathrm{B-V}} = 0.150$; age $=$ 1.0 Gyr; $\tau$ = 0.3 Gyr; Z $=$ 2.5\,Z$_{\odot}$;
log(M$_{*}$/M$_{\odot}) = $11.255.}
\label{all_4550}
\end{figure}

\begin{figure}
\epsscale{0.85}
 \plotone{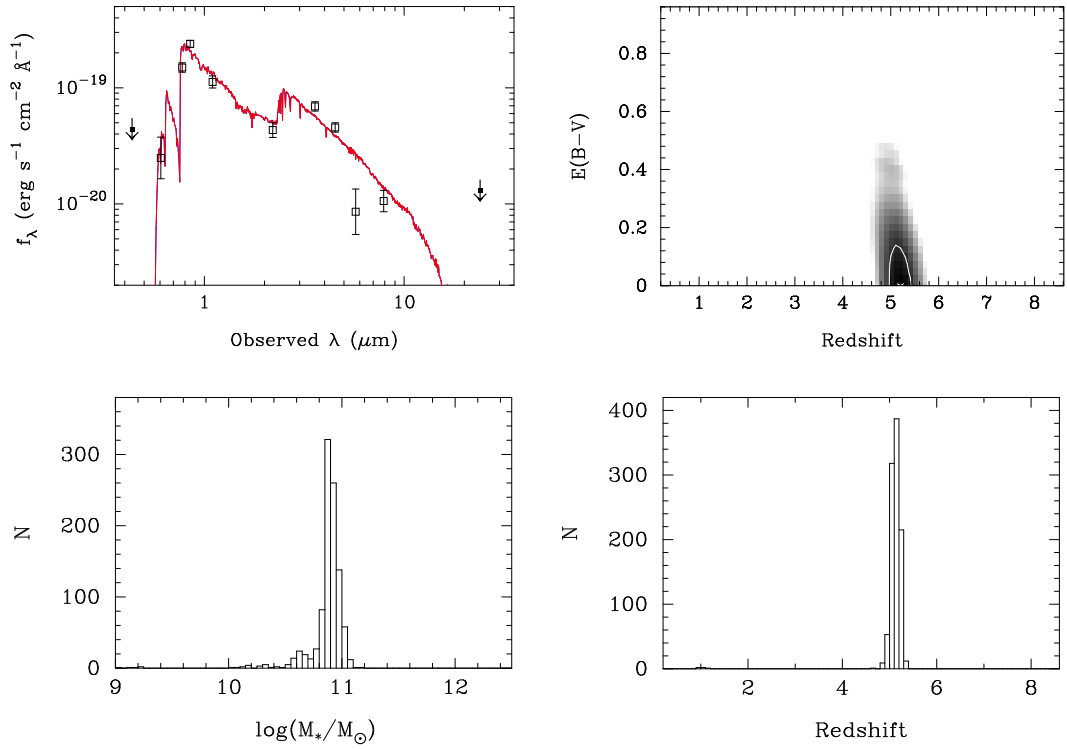}
\caption{Same as Fig.~\ref{all_0547} for BBG\#5197:
$z = 5.2$; E$_{\mathrm{B-V}} = 0.0$; age $=$ 0.9 Gyr; $\tau$ = 0.3 Gyr; Z $=$ 0.2\,Z$_{\odot}$;
log(M$_{*}$/M$_{\odot}) = $10.848.}
\label{all_5197}
\end{figure}

In Fig.~\ref{all_0547}$-$\ref{all_5197} we show the ACS (BV{\it i}{\it z}),
ISAAC (JHK$_s$) and Spitzer IRAC images of the 12 new $z \ga 5$ candidates
(for the corresponding plots for BBG\#3179/JD2, we refer to Mobasher et al. 2005).
The results from fitting BC03 models are also shown in Figs.~\ref{all_0547}$-$\ref{all_5197}.
The photometric data and the best--fit model SEDs from BC03 are shown
in the top left panel. The top right panel shows the distribution of
$\chi^{2}_{\nu}$ values as a function of extinction and redshift when
all other parameters are left free to vary at each (z, E$_{\mathrm{B-V}}$)
point. A wide spread of $\chi^{2}_{\nu}$ values is indicative of a degenerate
or unstable solution. Finally, the bottom left and right panels show the
result from $10^3$ Monte Carlo realizations for the distribution of
photometric redshift and stellar mass. Results for BBG\#3179 (HUDF--JD2)
can be found in Mobasher et al. (2005).

\subsection{Effect of Photometric Errors}\label{phot_error}

A major issue with the results obtained in this study is the presence
of large photometric errors in fluxes of individual galaxies, which could
significantly alter the shape of  their SEDs and their estimated parameters.
This becomes more serious as we combine observations from different
telescopes and instruments with very different characteristics. In this
sub--section we summarize the steps we take to verify the photometry
(cf. Sect.~\ref{data}) and to study the effect of photometric errors
on selection and photometry of our final BBG candidates.

The photometric errors affect our results in two ways. First, they could
lead to erroneous inclusion of objects into the BBG sample, or 
exclusion of some potential candidates. Second, they could affect the
observed SEDs and hence, the final sample and their estimated parameters. 
Since our technique mainly relies on the size of the Balmer break at
$5 \la z \la 7$, (i.e. the $K_s - m_{3.6}$ colors), an examination of
the ISAAC and Spitzer  photometry is crucial. 

The large size of the IRAC PSF could lead to blending in some of our
sources. The effect of this is to brighten the IRAC magnitudes, 
leading to redder $K_s - m_{3.6}$ colors and false inclusion of a
galaxy into the BBG sample. We attempt to correct for this using a
smaller (3$^{\prime\prime}$) aperture for sources with a nearest
neighbor in the K--band of $\leq$3$^{\prime\prime}$, and then using
the appropriate corrections to convert these to `total' magnitude
(see Sect~\ref{data}).
We then repeated IRAC photometry on our final BBG candidates by
simultaneously modeling the light distribution of the BBG and the
galaxies close to it, letting the positions of the galaxies be part
of the fit.
%%%
The IRAC PSF is slightly asymmetric and we used empirically derived
PSFs (from star) for each separate IRAC band and each observing epoch.
%%%
Application of the {\tt GALFIT}
routine succesfully separates the flux contributions of neighboring
sources and the Balmer--break candidate in all but one case (see below).
This procedure was carried out on all the candidates and for all the four
IRAC bands, resulting in a set of independent `total' magnitudes.
As the Spitzer observations were done in two different epochs, with
different PSFs, we measured the "unblended" magnitudes separately for
both epochs.
No attempt was made to subtract extended emission by fitting a S\'{e}rsic
profile. In most cases the BBG candidate and the surrounding neighbors are
small enough to allow a simple PSF fitting to estimate the total flux.

\begin{figure}
\epsscale{0.8}
\plotone{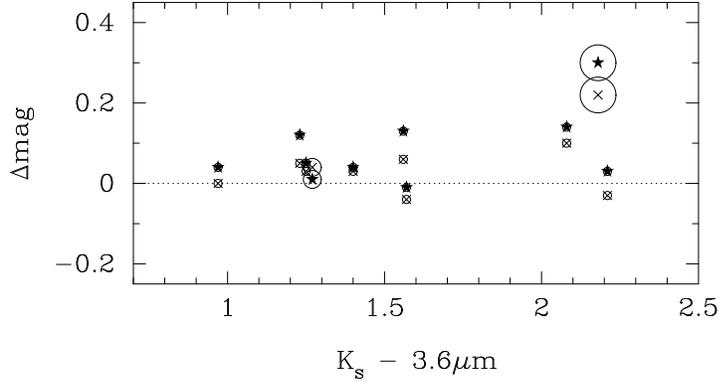}
\caption{
Comparison of IRAC magnitudes derived using aperture corrected m$_{\mathrm{total}}$ and those
derived using PSF fitting in {\tt GALFIT}. $\Delta$mag$ = m_{GALFIT} - m_{\mathrm{total}}$
is plotted against the K$_s - m_{3.6}$ colors where $m_{3.6}$ corresponds to m$_{\mathrm{total}}$.
The $\Delta$m are shown for 10 Balmer break galaxies (BBG\#3179/JD2 is not shown as its ISAAC and
IRAC photometry are obtained from Mobasher et al. 2005). Channel 1 (3.6$\mu$m) values are shown
as stars, channel 2 (4.5$\mu$m) values are shown as crosses.
Each data point is surrounded by a circle, where the size of the circle corresponds to the separation
between the coordinates of the ISAAC and IRAC centroids. Source with separations $\leq$0\ffas5 are
shown with small circles, those with separations in the range $0.5-0.9$ are shown with medium circles
and those with separations  $\geq$1\ffas0 with large circles. Only one BBG falls in the last category
(BBG\#2068).}
\label{mag_galfit}
\end{figure}

Fig.~\ref{mag_galfit} and Table~\ref{galfit} compares the IRAC `total'
magnitudes and those estimated using the {\tt GALFIT} routine. In general,
{\tt GALFIT} magnitudes are fainter. However, for the isolated sources,
the two magnitudes agree within 0.05 mag, giving support to our initial IRAC
photometry. For the sources were the magnitude difference is $\ga$0.1, the 
{\tt GALFIT} results suggests that blending is an issue (see Table~\ref{galfit}).
To explore the impact of this on the model fitting results, we re--fitted the
same BC03 models as before using the revised IRAC fluxes from {\tt GALFIT}.
The only significant change in the parameters defining the best--fit parameters
was found for (BBG\#4034), which now has a best--fit solutions as a dusty
starburst galaxy at $z = 5.5$, and is therefore disqualified as a Balmer--break
galaxy. For BBG\#4053, which has the largest correction to the IRAC magnitudes,
the {\tt GALFIT} results did not converge satisfactorily due to nearby extended
neighbors. Although its best--fit solution is still that of a BBG--type object,
we remove it from the sample as well. We discuss the individual galaxies
in Sect.~\ref{individual}.
Our final results for all Balmer--break candidates, including the Monte Carlo
simulations, are based on the magnitudes estimated using the {\tt GALFIT} procedure.

We examined the ISAAC near--IR magnitudes by performing multi-aperture 
photometry and estimating the `total' JHK magnitudes using 
individual growth curves. The effect of sky subtraction was examined
by performing different methods to independently measure and subtract the
background. The estimated `total' magnitudes from growth curves agree
well with MAG\_AUTO estimates, independently measured from SExtractor
(using background maps). We find an agreement better than 0.05 ($J$), 
0.07 ($H$) and 0.09 ($K_s$) mag. Furthermore, for the HUDF field 
(a sub--area of the GOODS--S), where both ISAAC and NICMOS JH magnitudes
are available, we compared these magnitudes for objects in common
(Mobasher \& Riess 2005) between the two instruments. The agreement
was $ < 0.05$ at the bright end and $< 0.10$ at the faint end. 

It is always possible that combined uncertainties in the ISAAC and IRAC
zero-points, or some other (presently unknown) effects in their
respective photometry, could lead to artificially red $K_s-m_{3.6}$
colors, and hence, erroneous selection of the BBGs or wrong estimates
of their physical parameters.  We investigate this in
Sect.~\ref{testing} by analyzing other, confirmed high--redshift galaxies in the
GOODS-S field, comparing their spectroscopic redshifts to photometric
redshifts derived in the same way as we have done for the BBGs, using Ê
ACS, ISAAC and IRAC photometry taken from the same catalogs

\subsection{Parameters of the candidates}\label{candidateparameters}

The final sample of Balmer break candidates with robust IRAC photometry
consists of 11 objects. ÊThe Êbest--fit parameter resulting from stellar population
model fits to the Êphotometry for these objects are presented in Table~\ref{table2}.
Here we also present parameter values calculated from the best--fit model
parameters: bolometric luminosity, stellar mass, the initial, current and average
star formation rates. As the star formation history is imposed by us and may not
reflect the actual chain of events, the values given in the table should be viewed
as indicative rather than absolute.

In Table~\ref{median}, we present the median values of parameters obtained
from the Monte Carlo simulations. The median values are, in most cases, not
very different from those obtained directly from the best--fit model fits using the
photometry given in Table~\ref{table2}, except that the median values for
metallicities tend to be higher. Overall, the metallicity is the least robust parameter
obtained from the model fits. In Table~\ref{median} we also list the percentage of
Monte Carlo realizations that result in a photometric redshift $z > 4$ and $z > 5$.
These values are indicative of the dispersion of the photometric redshifts
obtained from the Monte Carlo simulations, and hence, of the stability of the
solutions. In Table~\ref{table1} and~\ref{median} we have removed BBG\#4034
and BBG\#4053 due to potential blending in the IRAC bands
(see Sect.~\ref{phot_error}).

The $\chi^{2}_{\nu}$ value for the best--fit SED is $\la$2 for 10 of our 11 candidates.
The worst $\chi^{2}_{\nu}$ value (4.5) is found for BBG\#5197, which is the only galaxy
in our sample with a spectroscopically determined redshift
($z_{\mathrm{spec}} = 5.552$; $z_{\mathrm{phot}} = 5.2$; Vanzella et al. 2006).
%%%
The high  $\chi^2$ value is mainly due to one deviating IRAC photometric data point:
5.8$\mu$m appears to be too faint relative to the rest of the IRAC data points.
This cannot be caused by an emission line and the cause for the deviation remains
undetermined.
%%%
The confidence level in the redshift can also be evaluated from the distribution
of $\chi^{2}_{\nu}$ values as a function of redshift and E$_{\mathrm{B-V}}$ as
well as from the Monte Carlo simulations. A lower confidence of the photometric
redshift is indicated by: (1) uncertain photometry caused by source confusion in
the IRAC bands, (2) a large dispersion of the redshift distribution from the Monte Carlo
simulation, and, (3) the existence of a strong bi--modal solution (i.e. low--z, high
extinction vs. high--z, low extinction). 
In fact, for BBG\#5197, the Monte Carlo simulations indicate a stable photometric
redshift distribution despite having the worst $\chi^{2}_{\nu}$ value.
A possible reason for a high $\chi^{2}_{\nu}$ value is the presence of an AGN
component and strong emission lines, for which the current SED models are not
suitable. Since only one of our final candidates is detected in X--rays (BBG\#3348;
see below), this does not appear to be a major problem.

Inspection of the results from the Monte Carlo simulations
(Figs.~\ref{all_0547}$-$\ref{all_5197}) show that a generic
feature of the Balmer--break candidates is the presence of two local
minima: (1) $z \ga 5$ with little or no dust obscuration, and (2)
$z \approx 2$ with E$_{\mathrm{B-V}} \approx 0.5-0.9$. This reflects the
well--known degeneracy between age, extinction and redshift.
This degeneracy is discussed in Sect.~\ref{color} and illustrated
in Fig.~\ref{comp_sed}.
The secondary minima at lower redshift is usually interpreted as
a dusty starburst galaxy. However, a large part ($\sim$40\%) of the lower
redshift solutions corresponds to galaxies characterized by an old
stellar population (elliptical galaxy) with a considerable amount of
dust extinction, and not a dusty starburst galaxy per se.

The star formation history is modeled as exponentially declining with a time
constant $\tau$. Except for two objects, the candidate galaxies have a
current level of star formation activity that is $<$5\% of the peak star
formation rate. The candidates with the highest ongoing star formation rate
relative to the peak activity are: BBG\#2068 (29\%) and BBG\#4550 (12\%).
In Table~\ref{table2} we list the initial, current ($t = t_{\mathrm{SB}}$) and
average star formation rates for the candidate galaxies. For those cases that
are best fitted by an instantaneous star formation episode ($\tau = 0$) we
arbitrarily assumed that the initial star formation activity is spread over 100 Myr.
It is worth keeping in mind that the star formation rates discussed here are
dependent on the assumed parametrization of the star formation history.
The most robust estimate is the average rate, i.e. the stellar mass divided
by the age of the stars. The stellar ages range from $1 \times 10^8$ to
$1 \times 10^9$ years, with corresponding formation redshifts
$z_{\mathrm{form}} = 6 - 26$ (see Table~\ref{table2}).

%%%
The metallicity is not strongly constrained by the solutions. This is due
to the degeneracy between metallicity and age, as well as extinction. A
change in the metallicity can be offset by a small change in either age
and/or extinction. This is evident in the Monte Carlo simulations, where
metallicity rarely shows a strongly prefered value. We tried fitting
the model SED keeping the metallicity at Solar. The results were not
significantly different from when using all four metallicity values in the fit.

We measured the half-light radii, $r_{\mathrm{h}}$, of the Balmer--break
candidates by applying 16 apertures of increasing radius to each galaxy
and estimating the encircled flux. The smallest radii was of similar size
as the radius of point--spread function. We measured $r_{\mathrm{h}}$
on the K$_s$ images, obtained with VLT/ISAAC, with a PSF FWHM of $\sim$0\ffas4.
We derived the half--light radius in 13 of the BBG candidates (plus JD2).
In a few cases the growth curves did not turn over at the largest radii.
This is most likely due to blending with nearby sources.
All of the BBG's with measured $r_{\mathrm{h}}$ are resolved, with an average
half-light radius of $0\ffas34 \pm 0\ffas04$. However, since the light profiles
were not deconvolved with the PSF, they represent upper limits to the half--light
radius $r_{\mathrm{h}}$. At a redshift $z = 5.2$, this corresponds to a radius of
$\sim$2 kpc.

The GOODS--South field has been observed at radio wavelengths (1.4\,GHz)
to a limiting 1$\sigma$ sensitivity of 14$\mu$Jy using the Australian Compact
Telescope Array (Afonso et al. 2006).  A total of 64 radio sources are found within
the GOODS--South field, but none is associated with our Balmer break candidates.
Only one of the BBG's in our sample is detected in the 1\,Msec X--ray survey
of the GOODS South field done with the Chandra X-ray Observatory (Giacconi
et al. 2002). The galaxy, BBG\#3348 is undetected in both the soft (0.5$-$2 keV)
and hard (2$-$8 keV) bands, but is marginally detected when combining the
two bands. The flux is $9.6 \times 10^{-17}$ erg s$^{-1}$ cm$^{-2}$. At a
redshift $z_{\mathrm{phot}}=5.1$, this translates into $L_{\mathrm{X}} =
3 \times 10^{43}$ erg s$^{-1}$ ($7 \times 10^{9}$ L$_{\odot}$).
This is more than two order of magnitudes larger than the typical X--ray
luminosity of Lyman--break galaxies at $z\sim3$ and $z\sim4$ (Lehmer et al. 2005).
The X--ray luminosity of BBG\#3348 is only $\sim$0.2\% of the
bolometric luminosity derived from the UV-to-IR part of the SED.
While this X--ray luminosity is too low for a QSO, it is consistent
with the presence of a low--luminosity AGN. The observed X--ray flux
probes rest--frame energies in the range 3--49 keV, where attenuation
due to a large hydrogen column density should be a negligible effect.
The remainder of the Balmer-break galaxies are undetected with Chandra,
with typical 3$\sigma$ upper limits of $5.1 \times 10^{-17}$ and
$2.4 \times 10^{-16}$ erg s$^{-1}$ cm$^{-2}$ in the soft and
hard bands, respectively.

\subsection{MIPS 24$\mu$m detections}\label{mips}

The GOODS-South field has also been surveyed using the
Multiband Imaging Photometer for Spitzer (MIPS)
at 24$\mu$m (Dickinson et al. in prep.; Chary et al. in prep.).
For galaxies at redshifts $z \sim 2$, the MIPS 24$\mu$m band probes
a region in the mid-IR corresponding to
redshifted PAH emission. For galaxies in the redshift range $z \approx 5 - 7$, the
MIPS band covers restframe $3 - 4\mu$m where emission features from
PAH's are weaker. Furthermore, PAH emission is associated with
star formation activity as well as the presence of gas and dust,
and since the $z > 5$ models that fit the photometry in general have a
low level of ongoing star formation and dust, we do not expect to find
strong MIPS emission for the Balmer--break candidates.

The MIPS 24$\mu$m fluxes are given in Table~\ref{table1}. Among the
11 candidates, 7 show emission at 24$\mu$m. One of the MIPS detected
BBG's is also a weak X--ray source (BBG\#3348). The flux densities range
from 20 to 83$\mu$Jy, with a median value of 42$\mu$Jy.
The 24$\mu$m catalog (Chary et al. in prep.) fits point sources to the MIPS
image at prior positions defined by the IRAC catalog. Therefore, each IRAC
source has a 24$\mu$m flux measurement and uncertainty, even if there is
no significant detection. We therefore associate the K--detected galaxies and
MIPS 24$\mu$m sources based solely on positional coincidence, where we
assume that objects are associated if their centers are located within a radius
of 1\ffas0.

Applying the same selection criterion for the entire K$_s-$selected catalog,
and requiring that the sources are detected at 24$\mu$m with S/N$\geq$5,
we find that for K$_s$ magnitudes in the range 20--22, the MIPS 24$\mu$m
detection fraction is close to constant at $\sim$55\%, while for K$_s = 23.5$,
the detection fraction decreases to $\sim$30\%.
Hence, based on magnitude alone, we would expect to find approximately 3
Balmer break galaxies with MIPS 24$\mu$m detection with a S/N$\geq$5.
Instead we find $\sim$55\% of the BBG's detected at 24$\mu$m, the same as
the detection fraction of the brighter galaxies at K$_s \sim 20-22$.

\begin{figure}
\epsscale{0.7}
\plotone{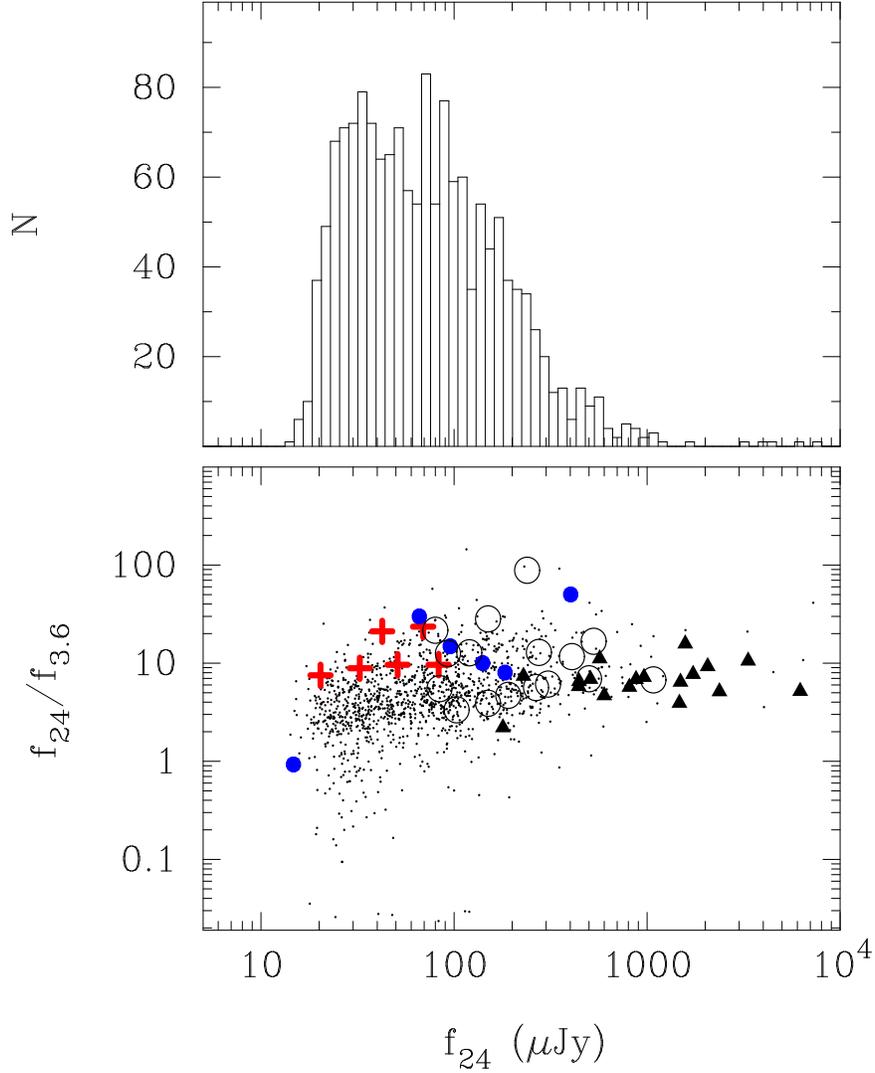}
\caption{The 24$\mu$m MIPS data for galaxies in the K--selected catalog, matched to the MIPS
catalog. The top figure shows the distribution of $24\mu$m fluxes ($\mu$Jy) for 1327 sources
with a positional coincidence with MIPS 24$\mu$m sources of $\leq$1\ffas0.
The bottom figure shows the flux density ratio $f_{24\mu\mathrm{m}}/f_{3.6\mu\mathrm{m}}$ 
vs. the $f_{24\mu\mathrm{m}}$ for the 1327 sources (small dots). In addition we also show
the 7 BBGs detected at 24$\mu$m (crosses),
a sample of $z\sim3$ Lyman--break galaxies (filled circles) (Rigopoulo et al. 2006),
AGNs at redshift $z\sim 4.5-6$ (filled triangles) (Hines et al. 2006; Jiang et al. 2006),
and submillimeter detected galaxies (open circles) (Ashby et al. 2006).
}
\label{mipsfig}
\end{figure}

In Fig.~\ref{mipsfig} we compare the observed 24$\mu$m properties of
the galaxies in the K--selected sample with those of the MIPS detected
Balmer--break candidates as well as a small sample of dusty star forming
galaxies at $z \approx 2-3$ (references in the figure caption).
Also shown in Fig.~\ref{mipsfig} are LBG's at $z\sim 3$, AGN's at $z \sim 4.5-6$,
and sub--millimeter detected galaxies at $z \sim 2$. While the Balmer
break galaxies are fainter than the comparison objects, their flux ratios
are roughly similar to both $z \sim 3$ star forming galaxies as well as $z \sim 5$
AGNs.
%%%
We also note that all MIPS detected BBG's have solutions which include a
substantial amount of internal extinction. In general the $E_{\mathrm{B-V}}
>0.2$ for the MIPS detected galaxies, while it is negligible for the non--MIPS
galaxies. There are two exceptions: BBG\#3348, which has a strong 24$\mu$m emission
but zero extinction in the best model fit, this particular galaxy is the
only one detected in X--rays, and BBG\#3179 (JD2), which is the BBG with
the highest redshift.

The high detection rate among the Balmer--break candidates
is surprising given the low levels of star formation activity
implied by the SED fits (cf. Table~\ref{table2}).
The galaxy HUDF--JD2 at $z_{\mathrm{phot}} \approx 6.5$ is also detected
at 24$\mu$m and in Mobasher et al. (2005) it was showed that this emission
can be consistent with the presence of an obscured AGN. The SED associated
with such an AGN has a minimal impact on the part of the SED covered by the
ACS/ISAAC/IRAC bands. While the presence of a super--massive black hole in
the Balmer--break galaxies would be expected if they follow the local
correlation between stellar and black hole mass (Magorrian et al. 1998;
Gebhardt et al. 2000), the prevalence of relatively strong 24$\mu$m
emission in the Balmer--break candidates
and the lack of X--ray emission
remains a challenge and concern for the BBGs.
In the following we will discuss two scenarios for the Balmer--break
galaxies: (1) all 11 candidates are considered as real, and (2) only
the 4 candidates without detectable 24$\mu$m emission will be considered as
likely candidates (BBG\#547, 3361, 4071, 5197). The latter defines the `no--MIPS' sample.
Two of the MIPS detected BBGs are only detected at the $\sim$5$\sigma$ level
(BBG\#2068 and BBG\#4550), but are considered as MIPS detected in the following.

\subsection{Individual sources}\label{individual}

\begin{itemize}

\item {\bf BBG\#0547}\ 
This is an isolated source
It is not detected at MIPS 24$\mu$m. The Monte Carlo simulations give
67\% probability for being at $z > 5$ (94\% for being at $z > 4$).
The SED fit is good, except that the K-band flux is off by a substantial
amount.

\item {\bf BBG\#2068}\ 
This source has a neighbor 2\ffas2 away. The two sources appear well
separated and the {\tt GALFIT} procedure converge satisfactorily.
The Monte Carlo simulations give $\sim$90\% probability for this source
to be at $z > 5$. The internal extinction is high, with E$_{\mathrm{B-V}} = 0.3$.
It is detected in the MIPS 24$\mu$m band with a S/N$\approx$5 and a
coordinate offset $\sim$1$^{\prime\prime}$. The MIPS detection is
therefore uncertain.

\item {\bf BBG\#2864}\ 
This is an isolated source, clearly seen in the IRAC bands and is detected
in the MIPS 24$\mu$m band. It is detected in the ISAAC K--band and
marginally in the H-band. It remains undetected in the
BV{\it iz} and J bands. The model SED represents an excellent fit,
with a substantial amount of internal extinction (E$_{\mathrm{B-V}}
= 0.25$). The Monte Carlo simulations, however, give a very broad
redshift distribution, with a 68\% probability of $z > 5$ and 77\%
probability for $z > 4$. The reason for the broad distribution is
that the source is poorly constrained due to the upper limits.

\item {\bf BBG\#2910}\ 
This is an isolated source, with the nearest neighbor at a distance of
3\ffas0. Nevertheless, since the neighbor is relatively bright foreground
object, {\tt GALFIT} photometry gives corrected IRAC magnitudes.
The Monte Carlo simulations imply a 16\% probability of $z > 5$
and 67\% for $z > 4$, consistent with the best-fit redshift of
$z = 4.9$ and a relatively narrow redshift distribution. The
source is detected in the MIPS 24$\mu$m band.

\item {\bf BBG\#3179}\ 
Detected by MIPS at 24$\mu$m. This object is also known as HUDF--JD2
and discussed in Mobasher et al. (2005). It is situated $\sim$7$^{\prime\prime}$
from a foreground galaxy. In the Hubble Ultra Deep Field, JD2 is undetected
in all 4 ACS bands at AB magnitudes $\ga$30 (see Table~\ref{table1}).
However, in our K--selected catalog, based on the shallower GOODS data,
it is flagged as a tentative detection in the V{\it iz} bands. Using this latter
photometry results in a photometric redshift $z = 5.1$, a stellar mass of
$5 \times 10^{11}$ M$_{\odot}$, and an age for the stellar population of
2 Gyr. This age is well in excess of the age of the universe and BBG\#3179
is the only candidate in our sample that violates the cosmological age restriction.
The SED fit has a $\chi^2 \approx 5$ and the Monte Carlo simulations give
a wide redshift distribution. Inspection of the UDF ACS/NICMOS images
reveals that BBG\#3179 is surrounded by several faint neigboring galaxies,
affecting the aperture based flux estimates. Careful subtraction of the neighbors
were done on the UDF data (Mobasher et al. 2005), leading to upper limits to
all four ACS bands and a photometric redshift $z = 6.5$.
While the stellar mass in this case is comparable to what we find using the
GOODS data, the age is found to be 0.6--1.0 Gyr. In the remainder of this work
we will base the JD2 results on the UDF data\footnote{The upper limits to the
ACS magnitudes used here are slightly modified compared to those presented
in Mobasher et al. (2005). In the latter analysis of the HUDF ACS data an aperture
of diameter 0\ffas48 was used to estimate the photometry, while an aperture of
0\ffas9 was assumed and quoted in the text. A re-analasys of the ACS data, using
the correct aperture diameter of 0\ffas9, masking the faint neighbors and
re-measuring the background noise amplitude, results in slightly brighter
limits to the ACS magnitudes of JD2. The corrected upper limits are given
in Table~\ref{table1}. These modified upper limits have no effect on the parameters
defining the best--fit SED or the stellar mass. However, for the Monte Carlo
simulations, the fraction of realizations with $z_{\mathrm{phot}} > 5$ decrease
from 85\% to 76\%. The NICMOS, ISAAC and IRAC photometry are not affected by this.}.
A recent analysis of this galaxy by Dunlop et al. (2007), gives the same
results as found in Mobasher et al. (2005) when using the same photometric data.
In addition, Dunlop et al. makes an independent assessment of the ACS data where
JD2 is assumed detected in the V{\it iz}--bands (albeit well below 3$\sigma$), at levels
$\sim$0.8 mag brighter than in Mobasher et al.'s analysis of the HUDF data. With
this set of ACS photometric data, Dunlop et al. find a best--fit photometric redshift
$z_{\mathrm{phot}} \sim 2.2$.

\item {\bf BBG\#3348}\ 
This is an isolated source, detected in the MIPS 24$\mu$m band. The
best--fit redshift is $z = 5.1$ and the Monte Carlo simulations give
a 33\% probability for $z >5$ and 67\% for $z > 4$, consistent with
the best--fit redshift. This is the only Balmer--break candidate
detected with Chandra. The X-ray luminosity is $3 \times 10^{43}$
erg s$^{-1}$, constituting only $\sim$0.2\% of the bolometric luminosity.
%This is a very good candidate.

\item {\bf BBG\#3361}\ 
An isolated source. No MIPS 24$\mu$m detection. The Monte Carlo simulations
give a 51\% probability for $z > 5$ and 100\% for $z > 4$, which
is consistent with the best-fit solution $z_{\mathrm{phot}}= 5.0$
and a small redshift dispersion.
%This is a very good candidate.

\item {\bf BBG\#4071}\ 
An isolated source, not detected in the MIPS 24$\mu$m band. The Monte
Carlo simulations give a redshift distribution with 51\% of the
realizations at $z > 5$ and 62\% at $z > 4$. The dispersion of redshifts
above $z = 4$ is, however, large, with median redshift of 5.1. The model
SED fits well, except for the H-band.

\item {\bf BBG\#4135}\ 
An isolated source with a detection in the MIPS 24$\mu$m band. The
redshift distribution from the Monte Carlo simulations is fairly broad,
with 35\% of the realizations at $z > 5$ and 71\% at $z > 4$, consistent
with the implied redshift $z = 4.9$. The model fits suggest a fairly
large amount of internal extinction (E$_{\mathrm{B-V}} = 0.35$).

\item {\bf BBG\#4550}\ 
There is a relatively bright neighbor at a distance of $\sim$2$^{\prime\prime}$.
BBG\#4550 is increasingly becoming brighter in the IRAC bands. The model SED
represents a fairly good fit, with some internal
extinction (E$_{\mathrm{B-V}} = 0.150$). The source is detected in the MIPS
24$\mu$m band. The Monte Carlo simulations give a redshift distribution with 22\%
at $z > 5$ and 79\% at $z > 4$. The relatively low probability for $z > 5$
is to be expected as the best-fit redshift is $z = 4.9$.

\item {\bf BBG\#5197}\ 
This source is spectroscopically confirmed at $z = 5.552$ (Vanzella et al. 2006).
The best--fit as well as the median photometric redshifts are both $z = 5.2$.
It is an isolated source with no MIPS 24$\mu$m detection. The Monte Carlo simulations
give a 93\% probability of the source being at $z > 5$ ($\sim$100\% for $z > 4$).
The model SED indicates no internal extinction. The fit, however, is not
perfect for the IRAC bands. This probably caused by a deviating $m_{5.8}$
photometric data point. The Monte Carlo simulations give a very narrow 
redshift distribution.

\end{itemize}

\section{The Balmer--Break Technique Applied to Other High--z Galaxy Samples}\label{testing}

The high-redshift Balmer--break candidates provide a formidable
challenge to spectroscopy. They are very faint at both
optical and near-infrared wavelengths and the
low level
of on-going
star formation means that strong emission lines, like Ly$\alpha$,
will be weak or non--existent. Nevertheless, one of the Balmer--break
candidates is spectroscopically confirmed.
We will discuss this particular galaxy in more detail below.

An alternate way to test the reliability of the parameters derived from
the SED fitting technique, in particular the photometric  redshift, is to
apply the models to galaxies with confirmed spectroscopic redshifts,
observed with the same, or similar, filter combinations as we use for
the BBG candidates in this paper. These galaxies will inevitably be
brighter and/or more actively star forming, but will provide a concrete
test of the procedures applied to the Balmer break sample.
We constructed a sample of galaxies with spectrsocopic redshifts from
several sources.

\begin{figure}
\epsscale{0.7}
 \plotone{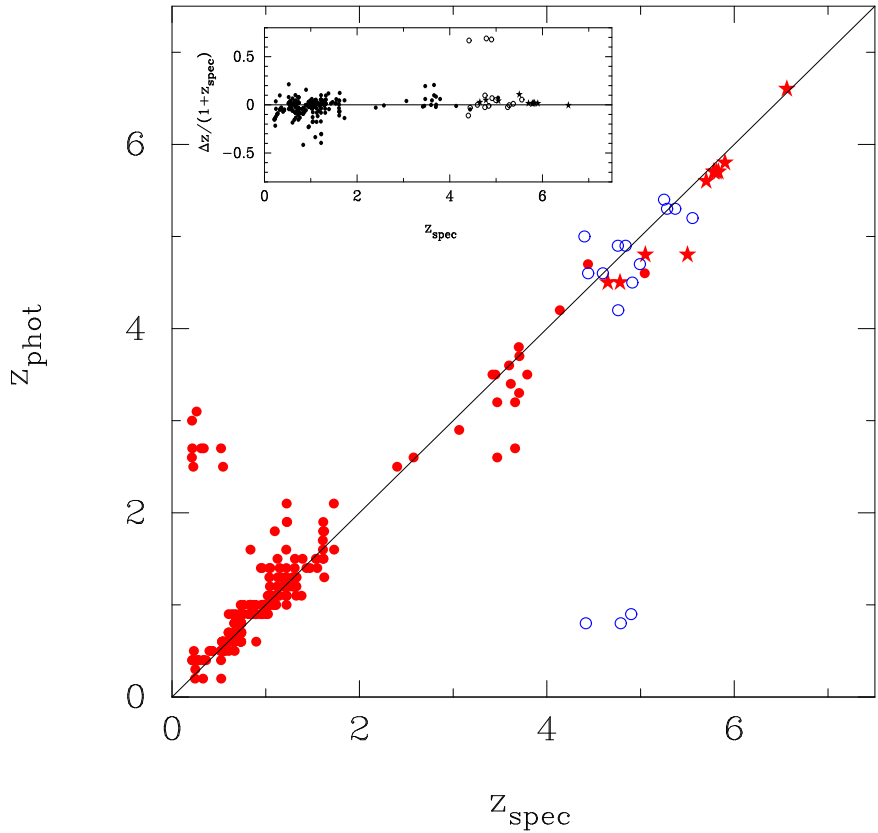}
\caption{A comparison between spectroscopic and photometric redshifts for galaxies in the
GOODS south field (394) shown as filled circles. The 25 additional high redshift objects for which
spectroscopic redshifts are known, and listed in Table~\ref{comparison}, are shown as stars.
The insert shows the distribution of $(z_{\mathrm{spec}} - z_{\mathrm{phot}}) / (1 + z_{\mathrm{spec}})$
as a function of redshift.}
\label{zcomp}
\end{figure}

The ESO/GOODS program of spectroscopy of galaxies in the GOODS--South field
has resulted in 807 optical spectra of 652 individual objects (Vanzella et al.
2006). From this survey we obtained 394 galaxies with securely determined
spectroscopic redshifts\footnote{We only retained galaxies with the highest
degree of confidence in the spectroscopic redshift determination in order to
not introduce extra uncertainties in the comparison with photometrically derived
redshifts.} These galaxies have redshifts in the range $z = 0.2 - 4.5$, and are
observed with the same telescope and filter combination as our BBG sample.
There is very good agreement between photometric and spectroscopic
redshifts in general (Fig.~\ref{zcomp}). An exception is a group of 9 outliers,
for which the photometric redshifts are higher than the spectroscopic ones.
The reason for this discrepancy is partly due to crowding, which makes
both spectroscopy and photometry difficult. In addition, most of the
deviant sources have a dual redshift solution, with a low redshift of slightly
higher $\chi^{2}_{\nu}$ than the high redshift solution. Since we do not apply
any priors to the photometric redshift determination, the solution with the lowest
$\chi^{2}_{\nu}$ is always chosen.
The scatter for $(z_{\mathrm{spec}} - z_{\mathrm{phot}})/(1+z_{\mathrm{spec}})$
has $\sigma = 0.12$ for all 394 galaxies.
Excluding the deviant sources, the scatter is  $\sigma = 0.06$. The inset in Fig.~\ref{zcomp}
shows the $(z_{\mathrm{spec}} - z_{\mathrm{phot}})/(1+z_{\mathrm{spec}})$ as a function
of redshift.

In addition to the Vanzella et al. (2006) galaxies, we also tested our fitting
technique on 23 high redshift sources with known spectroscopic redshifts
in the range $z_{\mathrm{spec}} = 4.4 - 6.6$ (Table~\ref{comparison}).
Yan et al. (2005) analyzed $i-$band drop-out galaxies in the HUDF,
six of which have spectroscopic redshifts ranging from
$z = 4.65$ to $z = 5.83$ (see Yan et al. 2005 for references to
spectroscopic measurements). One galaxy in the GOODS South field,
with spectroscopic redshift, was analyzed by Eyles et al. (2005). 
Both Yan et al. (2005) and Eyles et al. (2005) use broad--band photometry,
combining HST/ACS, HST/NICMOS and Spitzer/IRAC data. The photometric data
used in these studies are thus comparable to the photometric data used by
us for the entire GOODS--South field, with the exception that we use VLT/ISAAC
data for the JHK$_s$ bands.

We analyzed the spectroscopically confirmed galaxies from Yan et al.
(2005) and Eyles et al. (2005), using the same code as used for our
Balmer--break galaxies. As input we used the published photometry and
changed the filter response functions as necessary. We kept the redshift
as a free parameter, despite a known $z_{\mathrm{spec}}$.
The resulting photometric redshifts agree well with the spectroscopic
ones (see Table~\ref{comparison}), with differences typically less than
$|\Delta z| = |z_{\mathrm{spec}} - z_{\mathrm{phot}}| < 0.2$
(one exception is object \#15 in Yan et al., where
$z_{\mathrm{spec}} = 5.50$ and we obtained $z_{\mathrm{phot}} = 4.8$).

Both Yan et al. (2005) and Eyles et al. (2005) fitted Bruzual \& Charlot
(2003) model SEDs to some of their galaxies (Table~\ref{comparison}).
The best-fit models implies stellar masses of a few $\times 10^{10}$
M$_{\odot}$, ages of a few $10^8$ yrs with small amounts of extinction.
Metallicities, which is less well constrained than the other parameters,
is mostly consistent with solar metallicity. In our model fits, we find
parameter values which are very close to those obtained by both
Yan et al. (2005) and Eyles et al. (2005). The values derived by us
are compared to the published values in Table~\ref{comparison}.

We also fitted a model SED to the lensed galaxy HCM\,6A,
with a confirmed spectroscopic redshift $z = 6.56$ (Hu et al. 2002; see also
Chary et al. 2005; Schaerer \& Pell\'{o} 2005; Egami et al. 2005). Keeping the
redshift as a
free parameter, we derive a photometric redshift of $z = 6.6$ with a stellar
population age of $\sim300$ Myr. The stellar mass of HCM\,6A, corrected for
magnification due to lensing is $\sim 4 \times 10^{9}$ M$_{\odot}$. This is
an actively star forming galaxy and H$\alpha$ emission may introduce
flux in the 4.5$\mu$m IRAC band (e.g. Chary et al. 2005). While this may
introduce an error in the mass and age estimates, the photometric redshift
is not influenced.

Among the galaxies observed in the ESO/GOODS program of spectroscopy
of galaxies in the GOODS--South field (Vanzella et al. 2006) is the Balmer break
candidate BBG\#5197 (GDS\,J033218.92-275302.7).
A strong emission line with an asymmetric profile where the blue side is cut-off
is identified as Ly$\alpha$ at $z_{\mathrm{spec}} = 5.554$. Our photometric redshift
for this source is $z_{\mathrm{phot}} = 5.2$. This is the only source from our Balmer
break sample that is part of the ESO/GOODS spectroscopic survey.
A second line is seen at $\lambda_{\mathrm{obs}} = 972$nm, consistent with the
1483\AA\ NIV] line at the same redshift.  The NIV] emission line is not ordinarily
detected in star forming galaxies, but is seen in AGNs, where it is usually
accompanied by NV 1240\AA. Fosbury et al. (2003) discuss an object at
$z=3.36$ with similar UV emission lines which they interpret as a low--metallicity, Ê
low--mass, primeval galaxy with gas ionized by extremely hot, young stars.
It seems contradictory that a galaxy with an apparently massive, mature
stellar population implied by model fits to photometry for BBG\#5197 Ê
would have such properties, but conceivably the active star formation may Ê
apply to only a small fraction of the stellar mass, perhaps related to relatively
pristine gas in some component merging with the more massive, mature
host galaxy. ÊAlternatively, some unusual AGN may be reponsible for the
atypical UV emission line ratios, but this cannot easily explain the
apparent presence of a well-developed Balmer break discontinuity in the
SED of this galaxy.

Recently Stark et al. (2006) studied a sample of {\it z}--band selected galaxies
containing both star forming as well as quiescent galaxies, restricted to $z \sim 5$.
Spectroscopic redshifts for 14 of these galaxies had previously been obtained by
Vanzella et al. (2006) as part of a larger spectrscopic survey of the GOODS--South
field. The 14 galaxies have an average redshift $\bar{z} = 4.92$. One of the galaxies
observed by Vanzella et al., and included in the Stark et al. sample, is also part of
our Balmer--break sample, BBG\#5197 (ID GDS\,J033218.92-275302.7 in Vanzella et al.
2006; ID 32\_8020 in Stark et al. 2006).
Two additional galaxies with spectroscopic redshifts (ID 33\_10388 and 33\_10340)
have properties similar to the BBG galaxies (see Table~\ref{comparison}), but fall
just outside our color selection regions. This suggests that they are similar to our
post--starburst galaxies, but at a redshift $z < 5$. Indeed, these objects turn out to
have spectroscopic redshifts $z_{\mathrm{spec}} = 4.50$ and $z_{\mathrm{spec}} = 4.44$,
compared to our photometric redshifts of $z_{\mathrm{phot}} = 4.6$
and $z_{\mathrm{phot}} = 4.7$, respectively. These two galaxies are part of our K$_s$
selected sample and although they have properties consistent with them being massive
post--starburst systems,
they were excluded from our sample due to their lower redshifts. Nevertheless, the good
agreement between their spectroscopic and photometric redshifts lends support to the
derived photometric redshifts for the rest of the Balmer--break galaxies.

In a recent paper Dunlop et al. (2007) present a search for massive galaxies at
$z > 4$ using a version of the GOODS--South catalog similar to the one used here,
including Spitzer/IRAC data.
Instead of using a color selection, they fitted Bruzual \& Charlot (2003) models
to all galaxies in their sample and found 19 candidates with $z_{\mathrm{phot}} > 4$.
After further refinement of the photometry, they selected a final sample of 6 galaxies.
Their conclusion was that all of these were dusty and old galaxies at $z\sim 2$.
None of the objects in the Dunlop sample is part of our BBG sample, and it would be
interesting to explore the reason for this. The selection by Dunlop et al. was
done on objects with K$_{s} < 23.5$ (AB magnitudes). Only one of the original
13 BBGs in our sample have a K--magnitude as bright as this (BBG\#3348, the only
BBG with an X--ray detection). While this explains why our BBGs are not part of
Dunlop's sample, it does not exclude Dunlop's sources to be part of our sample.
All of Dunlop et al's galaxies are present in our K$_s-$selected catalog of 5754
galaxies. We identified Dunlop's 19 objects with implied $z_{\mathrm{phot}} > 4$
in our catalog and put them through
the same selection and fitting routine as for the BBGs. We immediately discard
4 of the 19 galaxies due to a positional offset between the ISAAC and IRAC
centroids in excess of 2$^{\prime\prime}$. Such a large offset strongly suggests
that the IRAC sources are blended with nearby neighbors and that the IRAC
photometry is unreliable. Of the remaining 15 galaxies, 9 objects fall
in our color selection regions (Sect.~\ref{color}), including five of the six
galaxies in Dunlop's revised sample.
Of the 9 galaxies fulfilling our color selection, the best--fit parameters
(obtained using the photometric values and errors from our catalog) shows that
7 have $z_{\mathrm{phot}} \la 2.5$. For the remaining 2 galaxies (Dunlop id: 2957
and 2958) we find $z_{\mathrm{phot}} = 4.6$ and 4.8, respectively. One of these
(id\#2957) has a very uncertain $z_{\mathrm{phot}}$, where the Monte Carlo simulations
give equal probability for a $z\approx 4.5$ and $z \approx 2.0$ solutions.
For the other source, id\#2958, our results agree well with that of Dunlop et al,
and it represents a good fit ($\chi^{2}_{\nu} = 1.03$). The implied stellar
mass is very large ($9 \times 10^{11}$ M$_{\odot}$). However, we do not find any
galaxy in Dunlop et al's sample that satisfy our redshift cut--off $z \geq 5$.

To summarize:
Our model fitting recovers known spectroscopic redshifts with a reasonably high
degree of accuracy for galaxies at both low and high redshifts. The parameters
derived by us for $z > 5$ galaxies agree with those obtained by other groups.
We interpret this as an indication that our technique is robust and reliable
when applied to the Balmer--break sample as a whole.

\section{Systematic Errors and Completeness}\label{errors}

\subsection{Effects of potential systematic errors}\label{systematics}

%Possible systematic errors
There are several sources of potential systematic errors that could affect our results.
Foremost is the reliability of the derived photometric redshifts. With only one
Balmer--break galaxy spectroscopically confirmed, we have to use indirect methods to assess
the confidence of the redshift estimates for the remainder of the candidates.
The photometric redshift technique was tested in the previous section (Sect.~\ref{testing}),
where we showed that the photometric redshifts obtained from our fitting technique are
robust, with an estimated success rate of $\sim$90\%.
Hence, assuming that the derived redshifts are approximately correct, we need to consider other
effects that could potentially lead to erroneous parameter values. Here we will be most
concerned with parameters essential for estimates of the stellar mass.

%{\bf Synthesis models.}\ 
In Mobasher et al. (2005) we used both the BC03 and Starburst99 (SB99; Leitherer et al. 1999;
V\'{a}squez \& Leitherer) models on the galaxy HUDF--JD2 and found them to give essentially
identical results when using the same parameterization of the star formation history.
However, in a recent paper, Maraston (2005) presented stellar synthesis models which
include a greater contribution to the red optical and near--infrared light from thermally
pulsating AGB stars for stellar populations of ages a few hundred Myr and older, compared
to models such as BC03 and SB99. For a given age, the Maraston models result a smaller
$M/L$  ratio than the BC03 and SB99 models. The difference becomes significant for
wavelengths $\ga$1$\mu$m.
The effect of the redder SED on fitting broad--band photometric data including near--infrared
and IRAC bands can be significant for redshifts $z < 4$, while for higher redshifts, the difference
in the model SEDs is most pronounced in the longest wavelength IRAC bands.
Nevertheless, the smaller $M/L$ ratio when using the Maraston models instead of BC03
or SB99 on stellar populations of ages from a few hundred Myr to $\sim$1 Gyr, i.e. the
range of ages considered for the BBG sample, could result in a lower estimated stellar
masses due to the increased near--infrared flux from AGB stars for a given stellar composition. 

%{\bf Initial mass function.}\ 
We use a Salpeter initial mass function with a lower and upper mass cut-off
at 0.1 M$_{\odot}$ and 100 M$_{\odot}$, respectively. Changing the lower
mass cut-off to 1 M$_{\odot}$, or using an IMF that is deficient in lower mass
stars relative to a Salpeter IMF, such as the IMF proposed by Chabrier
et al. (2003), would reduce the inferred stellar masses by a factor $\sim 1.5$.
It is, however, worth noting that the use of a Salpeter IMF with the upper and lower
mass cut--offs as used here, would not substantially change our masses relative to
that of other galaxies at similar or lower redshifts as long as the IMF remains
independent of redshift and galaxy mass.

It is presently difficult to estimate the combined effect of these potential
systematic errors. The effect of changing the IMF to a more top--heavy one,
and using the intrinsically redder SED from the Maraston models,
would both be to lower the average stellar masses of the Balmer break galaxies.
The magnitude of the effect is unknown, but could possibly be a factor $\sim$2
or more.

\subsection{Completeness}\label{completeness}

The K$_s$ selected sample is $\sim$82\% complete at $K_{\mathrm{AB}} \approx 23.5$
(Sect.~\ref{data}). The $K_{\mathrm{AB}}$ magnitudes for the BBG candidates range from 
$22.95 - 24.75$. The brightest candidate is the X--ray detected BBG\#3348, with the second
brightest BBG at $K_{\mathrm{AB}} = 23.84$. The average magnitude is  $K_{\mathrm{AB}}
= 24.2$. At this magnitude the completeness is $\sim$40\% (Fig.~\ref{maglimits}). This
represents the completeness, $\xi$, in selecting galaxies with K--magnitudes typical
for our selected BBGs.

\begin{figure}
\epsscale{0.9}
 \plotone{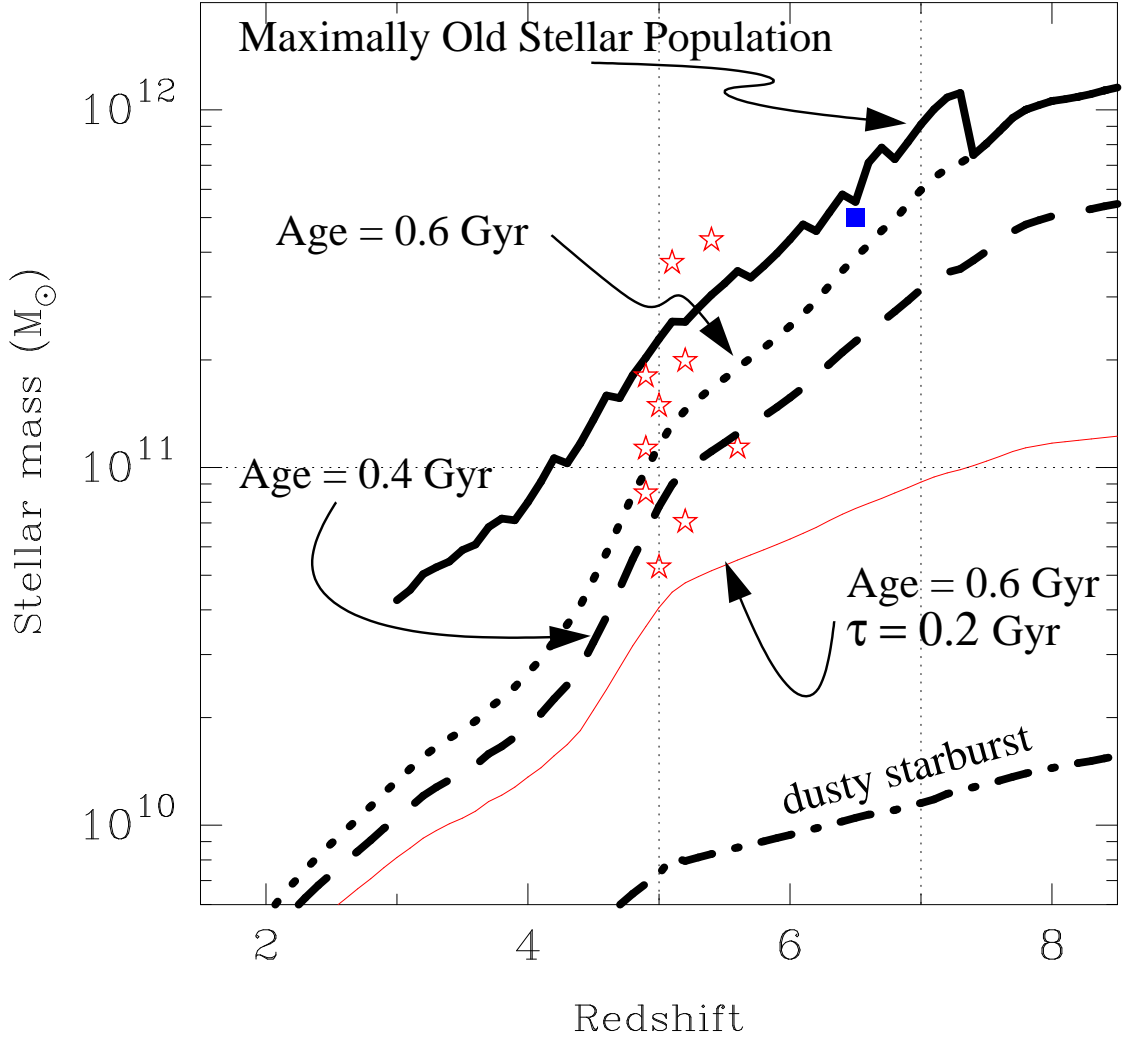}
\caption{Detection limits for model galaxies with an observed $K_{\mathrm{AB}} = 24.2$, as a function
of redshift. The thick black line represents a maximally old stellar population, i.e. a population
which is as old as the age of the universe at any given redshift. The wiggles on this line is due
to the discreteness of the age bins in the model SED (100 Myr). The maximally old population has a
solar metallicity, with all stars formed instantaneously (i.e. $\tau = 0.0$).
The dotted line corresponds to a passively evolving stellar population of a fixed age of 600 Myr and
with solar metallicity. The dashed line is for a fixed age of 400 Myr. The thin red line corresponds
to a stellar population where star formation started 600 Myr ago and the star formation
activity is declining exponentially with a $\tau = 200$ Myr.
For all these cases, E$_{\mathrm{B-V}} = 0$.
Finally, for comparison, the dash--dot line represents a starburst galaxy with some internal extinction
(E$_{\mathrm{B-V}}=0.15$, age$=$15 Myr, $\tau = 0$ and $Z_{\odot}$). The stars mark the
values derived for the BBG candidates (HUDF--JD2 is marked as a square).}
\label{masslimits}
\end{figure}

We also need to estimate the completeness in terms of stellar mass and age.
We do this by using a model SED with a fixed observed $K_s$ magnitude of 24.2
for all redshifts. We use a model with solar metallicity, no internal extinction
and characterized by instantaneous star formation, i.e. $\tau = 0$. Furthermore,
we use a maximally old stellar population, that is, at any given redshift, the
stellar population is assumed to be as old as the universe at that particular epoch
($z_{\mathrm{form}} = \infty$).
At $z = 5$, a maximally old stellar population has an age of 1.2 Gyr, and at
$z = 8.5$, the age is 0.6 Gyr.
The stellar mass is derived in the same manner as for the BBG candidates: the
M$_{*}$/L$_{\mathrm{bol}}$ ratio is obtained from the BC03 model given the age of
the stellar population, and the bolometric luminosity is obtained by integrating
over the SED, normalized to an observed $K_{\mathrm{AB}} = 24.2$. The resulting stellar
mass as a function of redshift is shown in Fig.~\ref{masslimits} as a thick black
line. The sawtooth appearance is due to the discrete age bins ($\Delta t = 100$ Myr)
used for the models. Any galaxy with a younger stellar population, a more
extended star formation history (i.e. $\tau > 0$), or lower metallicity, would have a
detection limit at a lower stellar mass. A higher metallicity and/or significant
internal extinction, on the other hand, would increase the stellar mass needed
for detection with $K_{\mathrm{AB}} = 24.2$.
In Fig.~\ref{masslimits} we also show the detection limits for a galaxy
with a fixed age of 600 Myr and $Z = Z_{\odot}$ (dotted line), a fixed age
of 400 Myr and $Z = Z_{\odot}$ (dashed line). The other parameters,
(E$_{\mathrm{B-V}} = 0.0$ and $\tau = 0$) are the same as for the maximally old
stellar population. A maximally old stellar population of solar metallicity needs
to have a mass $> 2 \times 10^{11}$ M$_{\odot}$ at $z = 5$ and $> 9 \times 10^{11}$
M$_{\odot}$ at $z = 7$ in order to be detected at K$_{\mathrm{AB}} = 24.2$.
The stellar mass needed for detection is lower if the age is less than the maximally
old stellar population as well as if some residual star formation is ongoing.
To illustrate the latter point we also show the detection limit for a stellar
population of age 600 Myr and with an exponentially declining star formation rate
with $\tau = 200$ Myr (thin red line). In this case a $10^{11}$ M$_{\odot}$ galaxy
can be detected out to $z \sim 7.0$.
The final conclusion from this exercise is that at $z \ga 5$ we can only detect
galaxies more massive than a few $10^{11}$ M$_{\odot}$ if the stellar population
is maximally old and passively evolving. For younger populations and if star
formation is still ongoing (albeit at a much reduced level), we are sensitive
to stellar masses from a few $10^{10}$ M$_{\odot}$ to $10^{11}$ M$_{\odot}$.
Our best--fit masses for the $z \approx 5$ BBGs (Table~\ref{table2}) are in the
range $\log(M_{*}/M_{\odot}) = 10.7 - 11.7$. The preceeding analysis shows that
at $\log(M_{*}/M_{\odot}) > 11.3$, even maximally old galaxies should have K$_s <
24.2$ and thus our photometric completeness estimate (40\%) should be reasonable.
However, we may be progressively more incomplete to old galaxies without ongoing
star formation at masses $<2 \times 10^{11}$ M$_{\odot}$.

Without knowledge of the intrinsic properties characterizing the Balmer--break
galaxies, it is difficult to define the volume over which we sample the galaxies
given our selection criteria. The lower limit is of course set by our imposed
selection of $z \geq 5$. For the upper redshift limit we use the mass limits
depicted in Fig.~\ref{masslimits} to make a reasonable estimate based on the derived
properties of the BBGs in our sample. As listed in Table~\ref{table2}, 6 of the 11
galaxies have a current SFR of at least a few M$_{\odot}$\,yr$^{-1}$, the average
age of the stellar population is $0.8 \pm 0.3$ Gyr and 8 have $\tau > 0$. This
suggests that we should use a less than maximally old stellar population, with
a small amount of ongoing star formation, in defining the upper redshift limit.
We therefore estimate the upper limit based on a stellar mass $\ga 10^{11}$
M$_{\odot}$, age 0.6 Gyr and a $\tau = 0.2$ Gyr, giving an ongoing SFR of
$\sim$10 M$_{\odot}$\,yr$^{-1}$ and an upper redshift limit of $z = 7$
(Fig~\ref{masslimits}). The comoving volume for the redshift
interval $z = 5 - 7$, over the 145 arcmin$^2$ spanned by the GOODS South field\footnote{This
represents the mean of the coverage of the $J-$, $H-$ and $K-$bands (156 and 124 arcmin$^2$,
respectively; see Sect.~\ref{data}. It is slightly smaller than the GOODS ACS field (160 arcmin$^2$).}
is $7.0 \times 10^{5}$ Mpc$^{3}$. The effective comoving volume can
be expressed as $V_{\mathrm{eff}} = 7.3 \times 10^{5}\,\xi\,\eta$ Mpc$^{3}$, where
$\xi$ is the completeness for detecting galaxies with observed $K_{\mathrm{AB}} = 24.2$,
estimated to $\xi = 0.4$, and $\eta$ represents the completeness when accounting
for galaxies that may have dropped out of the selection for other reasons. To estimate
a value for $\eta$ requires knowledge, or an educated guess, of the population of
massive and evolved galaxies at these redshifts. We will not attempt this estimate
here, but will indicate when our ignorance of the completeness correction may affect
the derived quantities.
The effective comoving volume is $V_{\mathrm{eff}} = 2.9 \times 10^{5}\,\eta$ Mpc$^{3}$.

Assuming that all of the BBG candidates have correct redshift estimates, the comoving
number density of massive and old galaxies at redshift $z = 5 - 7$ is
$3.9 \times 10^{-5}\,\eta^{-1}$ Mpc$^{-3}$. Adding the individual stellar masses,
the total stellar mass becomes $2.3 \times 10^{12}$ M$_{\odot}$, giving a stellar
mass density of $8 \times 10^{6}\,\eta^{-1}$ M$_{\odot}$\,Mpc$^{-3}$.
The corresponding values for the no--MIPS sample (see Sect.~\ref{mips}) are
$1.4 \times 10^{-5}\,\eta^{-1}$ Mpc$^{-3}$ and
$1.4 \times 10^{6}\,\eta^{-1}$ M$_{\odot}$\,Mpc$^{-3}$, respectively.

\begin{figure}
\epsscale{1.0}
 \plotone{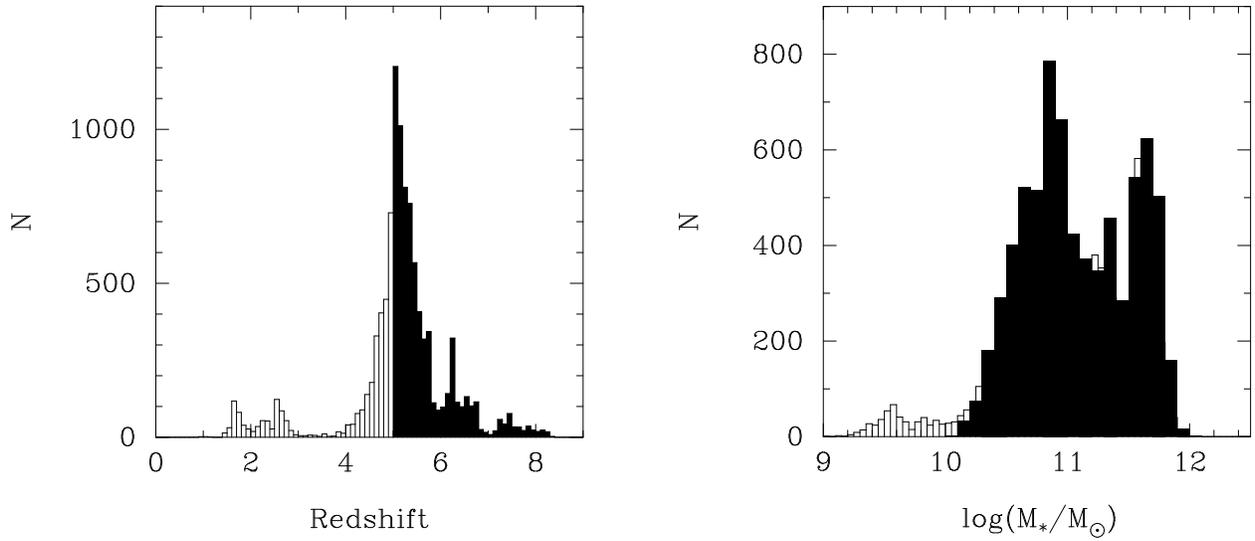}
\caption{{\bf Left:}\ The redshift probability distribution for all 11 Balmer--break candidates,
derived from the Monte Carlo simulations. The filled part corresponds to the solutions where
$z_{\mathrm{phot}} \geq 5.0$.
{\bf Right:}\ The probability distribution for the stellar mass. Again, the filled bars correspond
to the solutions for which $z_{\mathrm{phot}} \geq 5.0$.}
\label{all_hist}
\end{figure}

Our Monte Carlo simulations (Sect.~\ref{montecarlo}) allows us to estimate
the fraction of the BBGs that have photometric redshifts $z \ga 5$ when taking
the photometric errors into consideration. In Fig.~\ref{all_hist} we show the combined
probability distribution for photometric redshifts as well as the corresponding stellar
masses for the 11 Balmer--break candidates. Each of the the two distributions
contain $11 \times 10^3$ Monte Carlo realizations. The median redshift of the distribution
is $z_{\mathrm{med}} = 5.2$, the same as the average of the best--fit solutions. The
filled region corresponds to realizations with $z \geq 5$ and makes up 67\% of the
Monte Carlo realizations. This suggests that our initial selection criterion of $z \geq 5$
is fulfilled by $\sim 65-70$\% of our candidates, although from the simulation data we
cannot distinguish which ones.  Hence, our estimate of 11 BBGs with $z \ga 5$ needs
to be corrected for this, leading to an estimate of  $7 - 8$ BBG candidates.
The corresponding number density and stellar mass density should then be lowered
by a corresponding factor
($2.7 \times 10^{-5}\,\eta^{-1}$ Mpc$^{-3}$ and $5.4 \times 10^{6}$ M$_{\odot}$ Mpc$^{-3}$,
respectively).
The caveat with this analysis is the presence of Balmer--break galaxies at redshifts
$z_{\mathrm{phot}} < 5$, which may 'spill over' into the $z > 5$ range when the
photometric errors are considered. A preliminary analysis of the number of Balmer--break
candidates in the redshift range $4 \le z_{\mathrm{phot}} \le 5$, based on a
photometric redshift selection (Wiklind et al. in prep), suggests that the number
of BBG candidates at $z \ga 5$ is $\sim 10$, hence quite close to our initial 
estimate from the best--fit photometric redshifts. We therefore retain our number
of 11 BBGs in the $z = 5 - 7$ range.

\section{Discussion}\label{discussion}

The existence of massive and evolved galaxies at redshifts $z \ga 5$, when the universe
was $\la 1$ Gyr old, seems surprising at first sight. In the hierarchical scenario for
galaxy formation, the majority of massive galaxies are assembled at relatively low
redshifts. However, the presence of massive galaxies at high redshifts poses a fundamental
problem for hierarchical models only if their number density exceeds that of
correspondingly massive dark matter halos (e.g. Somerville 2004).
In Sect.~\ref{completeness} we derived the number and mass density of
the Balmer--break galaxies, using our sample of 11 galaxies, as well as for a
more restricted sample only containing those candidates which are not detected
with MIPS at 24$\mu$m, the 'no--MIPS sample'.

%Dark matter halos
By equating the co--moving number density of the Balmer--break galaxies with the
expected density of dark matter halos at the same redshift, we can estimate the
the maximum halo mass associated with the BBG's.
Using the Sheth--Tormen modified Press--Schechter formalism (Sheth \& Tormen 1999),
with the dark matter halo concentration predicted for the revised value of the power
spectrum normalization $\sigma_{8}=0.74$ (Spergel et al. 2007), and the  estimated
lower limit to the number density of BBGs ($3.9 \times 10^{-5}\,\eta^{-1}$ Mpc$^{-3}$),
we predict a halo mass of $M_{\mathrm{h}} = 1.0 \times 10^{12}$ M$_{\odot}$ (assuming
the standard $\Lambda$CDM model)\footnote{The estimated halo mass is a non-linear
function of the incompleteness coefficienct $\eta$. For instance, with $\eta = 0.5$,
the corresponding halo mass becomes $8 \times 10^{11}$ M$_{\odot}$.}.
Using the average stellar mass for the BBGs, we get $M_{*}/M_{\mathrm{h}} \approx 0.20$.
Considering the no--MIPS sample, with a number density $1.4 \times 10^{-5}\,\eta^{-1}$
Mpc$^{-3}$, the corresponding halo mass is $1.3 \times 10^{12}$ M$_{\odot}$, giving a
ratio $M_{*}/M_{\mathrm{h}} \approx 0.08$.
This estimate of the halo mass assumes that all available $\sim 10^{12}$ M$_{\odot}$
halos at $z \sim 5.2$ are associated with Balmer--break galaxies. If a fraction of
these halos would host lower mass stellar systems, such as Lyman--break galaxies,
the $M_{*}/M_{\mathrm{h}}$ ratio for the Balmer--break galaxies would have to
increase accordingly.

The ratio of the baryonic--to--total mass can be expressed in terms of a star formation
efficiency parameter, $\beta$ ($M_{*} = \beta\,M_{\mathrm{baryon}}$: the fraction of
baryons turned into stars over the life time of the galaxy), and the stellar mass,
M$_{*}$, as,
$$
M_{\mathrm{baryon}}/M_{\mathrm{total}} = \beta^{-1}M_{*}/M_{\mathrm{total}}
= \kappa
$$
where $M_{\mathrm{total}} = \beta^{-1}\,M_{*} + M_{\mathrm{h}}$. We can then write
$$
M_{*}/M_{\mathrm{h}} = \beta\,\kappa/(1 - \kappa).
$$
Adopting $\kappa = 0.17$ from the WMAP3 results (Spergel et al. 2007), we get
$M_{*}/M_{\mathrm{h}} = 0.20\,\beta$.
Klypin, Zhao \& Somerville (2002) estimate the total (virial) and baryonic mass of
the Milky Way and M31 galaxies and find $M_{*}/M_{\mathrm{h}} = 0.06 - 0.08$,
implying that in this case $\beta = 0.3 - 0.4$. For the Balmer--break galaxies, we
find $\beta \approx 0.4 - 1.0$, where the lower value corresponds to the no--MIPS
sample.
If we only consider the no--MIPS sample, the baryonic fraction is comparable to
local galaxies. However, for the full sample, the results suggests that
the BBGs at $z \approx 5.2$ contain a higher fractions of baryons than galaxies
at $z \approx 0$. Another possible explanation for the high baryonic fraction is that
the number density of dark matter halos at high redshift is underestimated by the
Sheth--Tormen analysis, or that we have systematically overestimated the stellar
masses of the BBGs by a factor $\ga 2$.
%%%
Using a Chabrier or Kroupa initial mass function with a less steep low--mass end,
could lower the estimated stellar masses by a factor 1.5--1.8 (see Sect.~\ref{systematics}).
%%%

It would be more instructive to compare the $M_{*}/M_{\mathrm{h}}$ ratio to
that of massive elliptical galaxies at $z \approx 0$. However, the evidence for
dark matter in elliptical galaxies is still circumstancial and limited to the central
regions. Using planetary nebulae and globular clusters as kinematic probes, it has
been possible to push the analysis to $\sim$5 $R_{\mathrm{eff}}$ (e.g. Romanowsky 2003;
Richtler 2004). While the number of ellipticals studied in detail is still small,
the general conclusion is that most have surprisingly weak dark matter halos,
i.e. large $M_{*}/M_{\mathrm{h}}$ ratios. It remains to be determined whether
the inferred $M_{*}/M_{\mathrm{h}}$ ratio for Balmer--break galaxies is consistent
with local giant elliptical galaxies.

%Stellar mass density
The stellar mass density of the universe from redshifts 0 to 6 has been estimated
by several groups, using different samples and methods (e.g. Bell et al. 2003; Dickinson
et al. 2003; Rudnick et al. 2003, 2006; Fontana et al. 2006; Yan et al. 2006). 
Some of these results are listed in Table~\ref{stellarmass}, as a comparison with the
results obtained for the BBGs. Most of the values listed in Table~\ref{stellarmass} are
based only on the objects observed and are lower limits. In a few cases, the mass function
has been integrated to obtain the total stellar mass (e.g. Dickinson et al. 2003).
In the local universe, the global stellar mass density is $(3-4) \times 10^{8}$\,M$_{\odot}$
Mpc$^{-3}$, while it decreases to $\sim 0.3 \times 10^{8}$ M$_{\odot}$ Mpc$^{-3}$ at $z = 2.5-3$.
In Sect.~\ref{completeness} we found that the stellar mass density of the 11 BBG candidates is
$8 \times 10^{6}\,\eta^{-1}$ M$_{\odot}$ Mpc$^{-3}$. This is $\sim 2-3$\% of the present
day total stellar mass density. Restricting the comparison to large early type galaxies in the
local universe, that is, galaxies at least as massive as our BBG sample, the percentage
increases to $\sim 4-6$\%. Comparing with the stellar mass density at $z \sim 2$, the
BBG sample already comprise $20-25$\% of the total stellar mass found at this redshift.
For the no--MIPS sample, the stellar mass density is $\sim$5 times smaller, and in this
case the comparison with stellar mass densities at lower redshifts has to be corrected
accordingly.

The galaxies found in this study are remarkable in that they contain a large
stellar mass, have small physical sizes and that their main epoch of star formation
occured at $z \ga 10$.
Galaxies with similar properties have, however, also been found by others.
In a recent paper, McClure et al. (2006) searched for Lyman--break galaxies in the
UKIDSS ultra deep survey, and found 9 candidates with $z > 5$. Their stellar masses
are $>5 \times 10^{10}$ M$_{\odot}$ and their ages range from 50--500 Myr. Overall,
these galaxies have properties similar to our Balmer--break galaxies. The number
density for the $z > 5$ galaxies found by McClure et al. is $\sim$4x smaller than
what we find in this paper. However, the different selection process, the fact that
the UKIDSS sample does not include IRAC data and the large completeness corrections
needed, makes a comparison difficult.
A number of massive galaxies at $z > 4$ were also found by Fontana et al. (2006) using
the GOODS-MUSIC sample. Their broad--band photometric data set consists of 14 bands,
including the 4 IRAC bands. The objects were identified by fitting template SED based on
Bruzual \& Charlot (2003) models to all galaxies in the sample.
The best-fitting SEDs for the high redshift objects suggest that they are passively evolving
galaxies, characterized by a very short time scale for star formation or by a constant star
formation and a large amount of dust extinction. The stellar masses found are in excess of
$10^{11}$ M$_{\odot}$.
Hence, massive and passively evolving galaxies at $z \sim 5$ are found in several studies.
A direct comparison of the results is presently not practical as different selection criteria
are used, and the completeness corrections are presently poorly defined.

Another surprising property of the Balmer--break galaxies is their compact sizes.
As derived in Sect.~\ref{candidateparameters}, the typical half--light radius is
$\la$2 kpc. Although this is larger than what is expected from the size vs. redshift
relation derived for UV bright galaxies at similarly high redshift (Ferguson et al.
2004; Bouwens et al. 2004; Dahlen et al. in prep.), the stellar masses of the Balmer--break
galaxies are at least 10 times higher. No massive compact galaxies of this type has been
found in the local universe.
However, compact galaxies with a large stellar mass have been found at $z \sim 1.4$
(Trujillo et al. 2006) and at $z\sim 2.5$ (Daddi et al. 2005b; Zirm et al. 2007; Toft
et al. 2007). These galaxies are massive ($M_{*} > 10^{11}$ M$_{\odot}$), with no sign
of AGN activity and contain a passively evolving stellar population, similar to the
Balmer--break galaxies. The effective radius of these galaxies, measured at rest--frame
optical wavelengths, are typically $\la 1$ kpc (Zirm et al. 2007; Toft et al. 2007),
or 3--6 times smaller than local counterparts of similar stellar mass. It is hypothesized
that the on--set of rapid star formation in these systems quench the star formation process,
leading to very compact objects. These galaxies, as well as the Balmer--break galaxies,
cannot represent fully assembled systems and must undergo subsequent evolution in their
structural parameters in order to resemble local galaxies with the same stellar mass.

%Implications for the reionization of the IGM
The presence of a population of massive galaxies that underwent
a period of intense star formation at $z \sim 10 - 25$ is likely to have
ramifications to the reionization of the intergalactic medium (IGM).
Panagia et al. (2005) calculated that the star formation associated
with the formation of the massive $z = 6.5$ galaxy HUDF--JD2 (Mobasher
et al. 2005), could significantly contribute to the reionization of the IGM.
The main uncertainties were the escape fraction of the Lyman continuum photons
and the volume density of objects similar to JD2.
With the new sample of post--starburst galaxies with formation redshifts
in the same range as JD2, it is possible to address this question.
The integrated output of Lyman continuum photons from the Balmer--break
candidates depends only on the total stellar mass and the assumed IMF
(Panagia et al. 2005). Because the average stellar mass for the BBG candidates
is about a factor 2 smaller than for JD2, assuming the same IMF, the average
number of Lyman continuum photons is likewise a factor of 2 lower.
Panagia et al. (2005) concluded that if each field of 2\ffam5 $\times$ 2\ffam5
contained a source like JD2, then these sources account for at least $\sim$20\%
of the reionization of the IGM. A higher percentage is possible if the escape
fraction is higher and/or the IGM is clumped. In the present case, we have a
total area that is 25 times larger and  a total ionizing photon output $\sim$10
times larger than in the case of JD2. Hence, the implication is that the BBG
sources can account for $\sim$10\% or more of the photons needed for reionization,
depending on the poorly constrained parameters describing the Lyman continuum
escape fraction and the clumpiness of the IGM itself. The implications for
reionization are discussed in more detail in Panagia et al. (in prep.).

\section{Summary}\label{summary}

In this paper we present evidence for a population of very massive and evolved
galaxies at $z \ga 5$. The results have been obtained by combining HST/ACS,
VLT/ISAAC and Spitzer/IRAC broadband photometric data on the GOODS southern
field.

Our main results are:

\begin{itemize}

\item Using the K$_s - 3.6\mu$m color index as the primary diagnostic for identifying
evolved stellar systems at $z \ga 5$, and using additional colors (both J$-$K and
H$-3.6\mu$m) to aid the separation of high redshift candidates from lower redshift
interlopers, we defined a sample of 134 potential candidates.
Fitting Bruzual \& Charlot (2003) models to the candidates, allowing for an extended
parameter space including redshift, age, internal extinction, metallicity and star formation
history, we extract 11 galaxies which are at redshift $z \ga 5$.
The confidence limits of the fitted parameters are tested through Monte Carlo simulations,
where the photometry is allowed to vary stochastically within their formal errors.
One of the candidate has a spectroscopically confirmed redshift agreeing with our photometric
estimate.

\item The 11 candidates have an average stellar mass of $2 \times 10^{11}$ M$_{\odot}$,
ages of 0.2$-$1.0 Gyr, and sub-solar to solar metallicities. Most of the candidates only
have small amounts of dust obscuration and low levels of ongoing star formation. 
One of the candidates is detected in X-rays with Chandra. The X--ray luminosity is
$\sim$0.2\% of the bolometric luminosity. The formation redshift of the candidate galaxies
range from $z_{\mathrm{form}} = 6$ to $z_{\mathrm{form}} \geq 25$.
The completeness of our sample is estimated to be $\sim$40\% based on the K$_s$ selection only.
However, we may be progressively more incomplete to old galaxies without ongoing
star formation at masses $<2 \times 10^{11}$ M$_{\odot}$.

\item Seven of the eleven BBG candidates are detected in the MIPS 24$\mu$m band, including
one X--ray detected source. The high detection rate is surprising given the large redshift
implied by the model fits and the low level of ongoing star formation.
%%%
While the 24$\mu$m detections could indicate PAH emission for galaxies at $z \sim 2-3$,
it is also consistent with a dust obscured AGN at $z \ga 5$.
We note that for the $z \ga 5$ solutions, six of the seven MIPS detected BBG's have
significant internal extinction, while the galaxies in the non--MIPS sample appear to
be essentially dust--free. The only exceptions are BBG\#3348, which is the only X--ray
detected BBG and BBG\#3179 (JD2) the highest redshift source.
%
%While the observed 24$\mu$m emission is consistent with an
%obscured AGN
The large number of MIPS detected sources is nevertheless surprising. We therefore also define
a smaller `no--MIPS' sample of BBGs, consisting of the 4 sources not detected with MIPS
at 24$\mu$m and derive number-- and stellar mass densities for both samples.

\item The comoving number density of $z \ga 5$ galaxies is $3.9 \times 10^{-5}\,\eta^{-1}$ Mpc$^{-3}$,
where $\eta$ represents unknown completeness corrections, when including all 11 candidates. For
the no--MIPS sample, the corresponding value is $1.4 \times 10^{-5}\,\eta^{-1}$ Mpc$^{-3}$.

\item The stellar mass density of galaxies more massive than $10^{11}$ M$_{\odot}$ at $z \approx 5.2$
is 2$-$3\% of the total stellar mass density at $z \approx 0$ and 20$-$25\% of the stellar mass density
at $z \sim 2$. For the no--MIPS sample, these values are smaller by a factor $\sim$2.2.

\end{itemize}

\acknowledgements
We thank Chien Peng for help with {\tt GALFIT}.
This paper is based on observations taken with the NASA/ESA {\it Hubble Space Telescope}, which
is operated by AURA Inc. under NASA contract NAS5-26555, {\it Spitzer Space Telescope}, and
the {\it Very Large Telescope} operated by the European Southern Observatory. Support for this work,
part of the Spitzer Space Telescope Legacy Science Program, was provided by NASA through
contract number 1224666 issued by the Jet Propulsion Laboratory, California Institute of Technology,
under NASA contract 1407.

\clearpage
\begin{deluxetable}{lccccc}
\tablewidth{0pt}

\tablecaption{Parameters for defining selection area in color-color diagrams\label{parameters}}

\tablehead{
\multicolumn{1}{c}{Model galaxy}                          &
\multicolumn{1}{c}{Age}                                   &
\multicolumn{1}{c}{E$_{\mathrm{B-V}}$}                    &
\multicolumn{1}{c}{$\tau$}                                &
\multicolumn{1}{c}{Z}                                     &
\multicolumn{1}{c}{Redshift range}                        \\
\multicolumn{1}{c}{}                                      &
\multicolumn{1}{c}{Gyr}                                   &
\multicolumn{1}{c}{}                                      &
\multicolumn{1}{c}{Gyr}                                   &
\multicolumn{1}{c}{Z$_{\odot}$}                           &
\multicolumn{1}{c}{}                                      \\
}

\startdata
Post-starburst  & $0.3 - 1.0$     &  $0.0 - 0.2$  &  0.0          &  $0.2 - 2.5$  & $1-8$ \\
Dusty starburst & $0.005 - 0.030$ &  $0.4 - 0.7$  & $0.0 - 0.2$   &  $0.2 - 2.5$  & $1-8$ \\
Elliptical      & $1.0 - 2.4$     &  0.0          &  0.0          &  $0.4 - 1.0$  & $1-4$ \\
\enddata
\end{deluxetable}

\clearpage

\begin{deluxetable}{ccccccccccccccc}
\tablewidth{0pt}
\tabletypesize{\scriptsize}
\rotate
\tablecaption{Photometric data for $z>5$ candidates\label{table1}}

\tablehead{
\colhead{ID}                       &
\colhead{RA}                       &
\colhead{DEC}                      &
\colhead{B}                        &
\colhead{V}                        &
\colhead{i}                        &
\colhead{z}                        &
\colhead{J}                        &
\colhead{H}                        &
\colhead{K$_s$}                    &
\colhead{3.6$\mu$m}                &
\colhead{4.5$\mu$m}                &
\colhead{5.7$\mu$m}                &
\colhead{8.0$\mu$m}                &
\colhead{24$\mu$m\tablenotemark{a}}          \\
\colhead{}                         &
\colhead{}                         &
\colhead{}                         &
\colhead{$\sigma_{\mathrm{B}}$}    &
\colhead{$\sigma_{\mathrm{V}}$}    &
\colhead{$\sigma_{\mathrm{i}}$}    &
\colhead{$\sigma_{\mathrm{z}}$}    &
\colhead{$\sigma_{\mathrm{J}}$}    &
\colhead{$\sigma_{\mathrm{H}}$}    &
\colhead{$\sigma_{\mathrm{K}_s}$}  &
\colhead{$\sigma_{\mathrm{3.6}}$}  &
\colhead{$\sigma_{\mathrm{4.5}}$}  &
\colhead{$\sigma_{\mathrm{5.7}}$}  &
\colhead{$\sigma_{\mathrm{8.0}}$}  &
\colhead{$\sigma_{\mathrm{24}}$}   
}

\startdata
%---------------------------------------------------------------------
 547  & 3:32:24.73 & -27:42:44.3 &
$>$27.80  & $>$27.80 & 27.23  & 25.67  &
$>$25.50  & 25.68  & 24.14  &
23.21  & 22.98  & 22.67  & 22.42 &
$<$24 \\
  &  &  &
           &             &  $\pm$0.84  &  $\pm$0.27 &
           & $\pm$0.3   &  $\pm$0.28  &
$\pm$0.10  &  $\pm$0.10  &  $\pm$0.16  &  $\pm$0.15 &
-- \\
%---------------------------------------------------------------------
2068\tablenotemark{c}  & 3:32:26.78  & -27:46:04.2 &
$>$27.8  & $>$27.8  &  27.02 & 25.92  &
$>$24.55 & 25.10    & 24.89  &
23.01  & 22.80  & 22.03  & 21.78 &
22: \\
  &  &  &
           &             &  $\pm$0.62  &  $\pm$0.29  &
           &  $\pm$0.43  &  $\pm$0.28  &
$\pm$0.10  &  $\pm$0.10  &  $\pm$0.10  &  $\pm$0.15 &
$\pm$4:\\
%---------------------------------------------------------------------
2864  & 3:32:53.25 & -27:47:51.6 &
$>$27.8  & $>$27.8  & $>$27.1  & $>$26.60  &
$>$24.55 & 25.82  & 24.68  &
22.49  & 22.02  & 21.75  & 21.60 &
32 \\
  &  &  &
           &             &             &             &
           &  $\pm$0.65  &  $\pm$0.30  &
$\pm$0.10  &  $\pm$0.10  &  $\pm$0.10  &  $\pm$0.15 &
$\pm$6 \\
%---------------------------------------------------------------------
2910  & 3:32:30.27 & -27:47:58.2 &
$>$27.80 & $>$27.80  & 26.47  & 26.01  &
24.84    & 24.84     & 23.84  &
22.73  & 22.43 & 22.27  & 22.45 &
69 \\
  &  &  &
           &             &  $\pm$0.48  &  $\pm$0.38 &
$\pm$0.36  & $\pm$0.46   &  $\pm$0.14  &
$\pm$0.10  &  $\pm$0.10  &  $\pm$0.10  &  $\pm$0.15 &
$\pm$5 \\
%---------------------------------------------------------------------
3179\tablenotemark{b}  & 3:32:38.74 & -27:48:39.9 &
$>$29.83 & $>$30.26  & $>$30.07  & $>$29.44  &
27.02  & 24.94 & 23.95  &
22.09  & 21.80 & 21.60  & 21.38 &
51 \\
  &  &  &
           &             &             &          &
$\pm$0.32  & $\pm$0.07   &  $\pm$0.13  &
$\pm$0.10  &  $\pm$0.10  &  $\pm$0.10  &  $\pm$0.15 &
$\pm$10 \\
%---------------------------------------------------------------------
3348  & 3:32:17.22 & -27:49:08.0 &
$>$27.80 & $>$27.80 & $>$27.10 & 25.42  &
24.45  & 23.83  & 22.92  &
21.56  & 21.23  & 21.24  & 21.42 &
83 \\
  &  &  &
           &           &            &  $\pm$0.24  &
$\pm$0.17  & $\pm$0.16 & $\pm$0.06  &
$\pm$0.10  & $\pm$0.10 &  $\pm$0.19 &  $\pm$0.15 &
$\pm$4 \\
%---------------------------------------------------------------------
3361  & 3:32:29.97 & -27:49:09.0 &
$>$27.80 &  27.95 & 26.33  & 25.81  &
$>$24.55 & 25.83  & 24.72  &
23.46    & 23.42  & 23.37  & 23.01 &
$<$27 \\
  &  &  &
           &  $\pm$0.65  &  $\pm$0.30  &  $\pm$0.23  &
           &  $\pm$0.63  &  $\pm$0.22  &
$\pm$0.10  &  $\pm$0.10  &  $\pm$0.18  &  $\pm$0.15 &
-- \\
%---------------------------------------------------------------------
4034  & 3:32:10.22 & -27:50:27.8 &
$>$27.80 & $>$27.80 & 26.88  & 25.23  &
24.57  & 24.04  & 24.01  &
22.94  & 23.04  & 22.98  & 23.20 &
$<$34\\
  &  &  &
           &            &  $\pm$0.84  &  $\pm$0.25  &
$\pm$ 0.18 & $\pm$ 0.22 &  $\pm$ 0.17 &
$\pm$0.10  & $\pm$0.10  &  $\pm$0.16  &  $\pm$0.19 &
-- \\
%---------------------------------------------------------------------
4053\tablenotemark{d}  & 3:32:33.48 & -27:50:30.0 &
$>$27.80 & $>$27.80 & 25.99  & 25.44  &
$>$25.5  & 25.48  & 24.97  &
23.73  & 23.93  & 23.78  & 23.36 &
-- \\
  &  &  &
      &             &  $\pm$0.21  &  $\pm$0.17  &
      & $\pm$ 0.54  & $\pm$ 0.33  &
$\pm$0.10  & $\pm$0.10  &  $\pm$0.19  &  $\pm$0.26 &
-- \\
%---------------------------------------------------------------------
4071  & 3:32:27.07 & -27:50:31.4 &
$>$27.80  & $>$27.80 & $>$27.10  & 27.21  &
25.58  & 24.60  & 24.06  &
22.63  & 22.31  & 22.25  & 22.11 &
$<$26 \\
  &  &  &
                     &    &    &   $\pm$0.92 &
$\pm$0.35  &  $\pm$0.24  &  $\pm$0.14  &
$\pm$0.10  &  $\pm$0.10  &  $\pm$0.10  &  $\pm$0.15 &
--\\
%---------------------------------------------------------------------
4135  & 3:32:48.43 & -27:50:39.0 &
$>$27.80  & $>$27.80 & $>$27.1  & 25.80  &
25.90  & $-$    & 24.34  &
23.14  & 22.74  & 22.38  & 22.24 &
42 \\
  &  &  &
           &             &             &  $\pm$0.34 &
$\pm$0.46  & $-$         &  $\pm$0.27  &
$\pm$0.10  &  $\pm$0.10  &  $\pm$0.16  &  $\pm$0.15 &
$\pm$3 \\
%---------------------------------------------------------------------
4550\tablenotemark{c}  & 3:32:24.62 & -27:51:38.2 &
$>$27.80 & 27.35  & 26.05  & 26.42  &
$>$25.50 & $-$    & 24.75  &
22.82  & 22.58    & 22.25  & 21.67 &
20: \\
  &  &  &
           &  $\pm$0.44  &  $\pm$0.27  &  $\pm$0.54 &
           & $-$         &  $\pm$0.27  &
$\pm$0.10  &  $\pm$0.10  &  $\pm$0.10  &  $\pm$0.15 &
$\pm$4 \\
%---------------------------------------------------------------------
5197  & 3:32:18.91 & -27:53:02.5 &
$>$27.80  & 27.70  & 25.22  & 24.51  &
24.77     & 24.68    & 24.30  &
22.72  & 22.64  & 23.44  & 23.06 &
$<$25 \\
  &  &  &
           &  $\pm$0.45  &  $\pm$0.10  &  $\pm$0.08 &
$\pm$0.13  & $\pm$0.20   &  $\pm$0.16  &
$\pm$0.10  & $\pm$0.10   &  $\pm$0.49  &  $\pm$0.23 &
-- \\
\tableline
%---------------------------------------------------------------------
\enddata
\tablenotetext{a}{MIPS 24$\mu$m flux density in $\mu$Jy. All other entries in the
Table are AB magnitudes: $m_{\mathrm{AB}} = -2.5\,\log(f_{\nu}) + 8.90$.
Upper limits are 5$\sigma$.}
\tablenotetext{b}{BBG\#3179 $=$ JD2. The magnitudes given here are updated from the
Mobasher et al. (2005) HUDF values (Sect.~\ref{individual}). The GOODS
data directly from the K--selected catalog are (BV{\it iz}): $>$27.8, $26.83
\pm 0.26$, $26.58 \pm 0.43$, $26.00 \pm 0.30$; (JHK): $>$24.5, $24.24
\pm 0.28$, $24.29 \pm 0.25$; (IRAC ch1, ch2, ch3, ch4): $22.09 \pm 0.10$,
$21.80 \pm 0.10$, $21.60 \pm 0.10$, $21.38 \pm 0.15$.}
\tablenotetext{c}{Marginal MIPS 24$\mu$m detection: flux $\sim$ 5$\sigma$.}
\tablenotetext{d}{None--detection in MIPS 24$\mu$m, but lacks upper limit.}
\end{deluxetable}

\clearpage

\clearpage
\begin{deluxetable}{cccccccccccccccc}
\tablewidth{0pt}
\tabletypesize{\scriptsize}

\rotate

\tablecaption{Comparison of m$_{\mathrm{`total'}}$ and GALFIT IRAC magnitudes\tablenotemark{a} \label{galfit}}

\tablehead{
\multicolumn{1}{c}{ID}                                    &
\multicolumn{3}{c}{3.6$\mu$m}                             &
\multicolumn{1}{c}{ }                                     &
\multicolumn{3}{c}{4.5$\mu$m}                             &
\multicolumn{1}{c}{\ }                                    &
\multicolumn{3}{c}{5.8$\mu$m}                             &
\multicolumn{1}{c}{\ }                                    &
\multicolumn{3}{c}{8.0$\mu$m}                             \\
\multicolumn{1}{c}{}                                      &
\multicolumn{1}{c}{m$_{\mathrm{`total'}}$}                &
\multicolumn{1}{c}{m$_{\mathrm{GALFIT}}$}                 &
\multicolumn{1}{c}{$\Delta$mag\tablenotemark{b}}                    &
\multicolumn{1}{c}{\ }                                    &
\multicolumn{1}{c}{m$_{\mathrm{`total'}}$}                &
\multicolumn{1}{c}{m$_{\mathrm{GALFIT}}$}                 &
\multicolumn{1}{c}{$\Delta$mag\tablenotemark{a}}                    &
\multicolumn{1}{c}{\ }                                    &
\multicolumn{1}{c}{m$_{\mathrm{`total'}}$}                &
\multicolumn{1}{c}{m$_{\mathrm{GALFIT}}$}                 &
\multicolumn{1}{c}{$\Delta$mag\tablenotemark{a}}                    &
\multicolumn{1}{c}{\ }                                    &
\multicolumn{1}{c}{m$_{\mathrm{`total'}}$}                &
\multicolumn{1}{c}{m$_{\mathrm{GALFIT}}$}                 &
\multicolumn{1}{c}{$\Delta$mag\tablenotemark{a}}                    }

\startdata
   547 & 23.17 & 23.21 &  0.04 &\ & 22.98 & 22.98 &  0.00 &\ & 22.76 & 22.69 & -0.07 &\ & 22.18 & 22.42 &  0.24 \\
  2068 & 22.71 & 23.01 &  0.30 &\ & 22.58 & 22.80 &  0.22 &\ & 22.02 & 22.03 &  0.01 &\ & 21.88 & 21.78 & -0.09 \\
  2864 & 22.47 & 22.49 &  0.03 &\ & 22.06 & 22.02 & -0.03 &\ & 21.70 & 21.75 &  0.05 &\ & 21.55 & 21.60 &  0.05 \\
  2910 & 22.61 & 22.73 &  0.12 &\ & 22.38 & 22.43 &  0.05 &\ & 22.15 & 22.27 &  0.12 &\ & 22.24 & 22.45 &  0.21 \\
  3348 & 21.52 & 21.56 &  0.04 &\ & 21.20 & 21.23 &  0.03 &\ & 21.19 & 21.24 &  0.05 &\ & 21.29 & 21.42 &  0.13 \\
  3361 & 23.45 & 23.46 &  0.01 &\ & 23.38 & 23.42 &  0.04 &\ & 23.28 & 23.37 &  0.09 &\ & 22.95 & 23.01 &  0.06 \\
  4034\tablenotemark{c} & 22.72 & 22.94 &  0.22 &\ & 22.75 & 23.04 &  0.29 &\ & 22.79 & 22.98 &  0.19 &\ & 22.84 & 23.20 &  0.36 \\
  4053\tablenotemark{c} & 23.22 & 23.73 &  0.51 &\ & 23.33 & 23.93 &  0.60 &\ & 23.09 & 23.78 &  0.69 &\ & 23.26 & 23.36 &  0.10 \\
  4071 & 22.50 & 22.63 &  0.13 &\ & 22.25 & 22.31 &  0.06 &\ & 22.18 & 22.25 &  0.07 &\ & 22.01 & 22.11 &  0.10 \\
  4135 & 23.09 & 23.14 &  0.05 &\ & 22.71 & 22.74 &  0.03 &\ & 22.34 & 22.38 &  0.03 &\ & 22.22 & 22.24 &  0.02 \\
  4550 & 22.67 & 22.82 &  0.14 &\ & 22.48 & 22.58 &  0.10 &\ & 22.02 & 22.25 &  0.23 &\ & 21.46 & 21.67 &  0.21 \\
  5197 & 22.73 & 22.72 & -0.01 &\ & 22.68 & 22.64 & -0.04 &\ & 23.98 & 23.44 & -0.54 &\ & 23.05 & 23.06 &  0.01 \\
\enddata
\tablenotetext{a}{BBG\#3179 (HUDF--JD2) was not included here. IRAC photometry was taken from Mobasher et al. (2005)}
\tablenotetext{b}{$\Delta$mag : m$_{\mathrm{GALFIT}} - {m_{\mathrm{`total'}}}$}
\tablenotetext{c}{Removed from the sample due to large corrections to the IRAC magnitudes (see Sect.~\ref{phot_error}).}
\end{deluxetable}

\clearpage

\clearpage
\begin{deluxetable}{cccccccccccccccc}
\tablewidth{0pt}
\tabletypesize{\scriptsize}

\rotate

\tablecaption{Best--fit parameters for the final sample of $z>5$ Balmer--break candidates\label{table2}}

\tablehead{
\multicolumn{1}{c}{ID}                                    &
\multicolumn{1}{c}{$z$}                                   &
\multicolumn{1}{c}{E$_{\mathrm{B-V}}$}                    &
\multicolumn{1}{c}{t$_{\mathrm{sb}}$}                     &
\multicolumn{1}{c}{$\tau$}                                &
\multicolumn{1}{c}{Z}                                     &
\multicolumn{1}{c}{$\log{L_{\mathrm{bol}}}$}              &
\multicolumn{1}{c}{$\log{M_*}$}                           &
\multicolumn{3}{c}{SFR (M$_{\odot}$\,yr$^{-1}$)}          &
\multicolumn{1}{c}{$z_{\mathrm{form}}$}                   &
\multicolumn{1}{c}{r$_{\mathrm{e}}$}                      &
\multicolumn{1}{c}{MIPS}                                  &
\multicolumn{1}{c}{r$_{\mathrm{off}}$\tablenotemark{a}}              &
\multicolumn{1}{c}{$\chi^{2}_{\nu}$}                      \\
\multicolumn{3}{c}{}                                      &
\multicolumn{1}{c}{Gyr}                                   &
\multicolumn{1}{c}{Gyr}                                   &
\multicolumn{1}{c}{Z$_{\odot}$}                           &
\multicolumn{1}{c}{L$_{\odot}$}                           &
\multicolumn{1}{c}{M$_{\odot}$}                           &
\multicolumn{1}{c}{$t = 0$}                               &
\multicolumn{1}{c}{$t = t_{\mathrm{SB}}$}                 &
\multicolumn{1}{c}{average}                               &
\multicolumn{1}{c}{}                                      &
\multicolumn{1}{c}{arcsec}                                &
\multicolumn{1}{c}{detection}                             &
\multicolumn{1}{c}{arcsec}                                &
\multicolumn{1}{c}{}                                      }

\startdata
0547        &  5.6  &  0.025  &  0.8   &  0.2  &  2.5  &  11.5450  &  11.0561  &  580     &   11    &  142     &    17  & 0.32 & no  & 0.2 &  1.22   \\
2068        &  5.2  &  0.300  &  1.0   &  0.8  &  2.5  &  12.0919  &  11.2984  &  348     &  100    &  199     &    26  & 0.31 & yes & 1.0 &  2.00   \\
2864        &  5.4  &  0.250  &  0.9   &  0.1  &  0.2  &  12.0222  &  11.6361  & 4327     & $\sim$0 &  481     &    21  & --   & yes & 0.2 &  0.38   \\
2910        &  4.9  &  0.100  &  0.4   &  0.0  &  0.2  &  11.6519  &  11.0530  & $>$5000  &    0    &  282     &     7  & --   & yes & 0.2 &  1.09   \\
3179\tablenotemark{a} &  6.5  &  0.000  &  1.0   &  0.0  &  0.2  &  12.0000  &  11.6990  & $>$5000  &    0    &  500     & $>$35  & --   & yes & 0.1 &  1.90   \\
3348        &  5.1  &  0.000  &  0.9   &  0.1  &  0.2  &  11.9587  &  11.5726  & 3738     & $\sim$0 &  415     &    16  & 0.35 & yes & 0.1 &  1.22   \\
3361        &  5.0  &  0.000  &  0.8   &  0.2  &  1.0  &  11.2232  &  10.7209  &  268     &    5    &  263     &    12  & 0.32 & no  & 0.6 &  0.91   \\
4071        &  5.0  &  0.075  &  0.4   &  0.0  &  1.0  &  11.6491  &  11.1723  & $>$5000  &    0    &  372     &     7  & 0.30 & no  & 0.1 &  1.40   \\
4135        &  4.9  &  0.350  &  0.3   &  0.1  &  0.2  &  12.0515  &  10.9282  &  892     &   44    &  283     &     6  & 0.39 & yes & 0.2 &  0.68   \\
4550        &  4.9  &  0.150  &  1.0   &  0.3  &  2.5  &  11.7385  &  11.2552  &  622     &   76    &  178     &    18  & 0.40 & yes & 0.1 &  2.16   \\
5197        &  5.2  &  0.000  &  0.9   &  0.3  &  0.2  &  11.5282  &  10.8483  &  247     &   12    &   78     &    17  & 0.31 & no  & 0.1 &  4.39   \\
\enddata
\tablenotetext{a}{The JD2 (BBG\#3179) results are based on photometric data from the Hubble UDF (see Sect.~\ref{individual};
Mobasher et al. 2005)}
\end{deluxetable}

\clearpage

\clearpage
\begin{deluxetable}{ccccccccccl}
\tablewidth{0pt}
\tabletypesize{\scriptsize}

\rotate
\tablecaption{Median values from Monte Carlo simulations for the final sample of $z>5$ Balmer--break candidates\label{median}}

\tablehead{
\multicolumn{1}{c}{ID}                                   &
\multicolumn{1}{c}{$z$}                                  &
\multicolumn{2}{c}{Percentage$^{(a)}$}                   &
\multicolumn{1}{c}{E$_{\mathrm{B-V}}$}                   &
\multicolumn{1}{c}{t$_{\mathrm{sb}}$}                    &
\multicolumn{1}{c}{$\tau$}                               &
\multicolumn{1}{c}{Z}                                    &
\multicolumn{1}{c}{$\log{M_*}$}                          &
\multicolumn{1}{c}{$z_{\mathrm{form}}$}                  &
\multicolumn{1}{l}{Comments}                             \\
\multicolumn{2}{c}{}                                     &
\multicolumn{1}{c}{$z_{\mathrm{phot}} > 4$}              &
\multicolumn{1}{c}{$z_{\mathrm{phot}} > 5$}              &
\multicolumn{1}{c}{}                                     &
\multicolumn{1}{c}{Gyr}                                  &
\multicolumn{1}{c}{Gyr}                                  &
\multicolumn{1}{c}{Z$_{\odot}$}                          &
\multicolumn{1}{c}{M$_{\odot}$}                          &
\multicolumn{1}{c}{}                                     &
\multicolumn{1}{l}{}                                                        
}

\startdata
0547        &  5.4  &  95.1   &  67.3   & 0.100  &  0.8   &  0.4  &  2.5  &  11.010  &    15 &  \\
2068        &  5.2  &  98.0   &  89.8   & 0.325  &  0.9   &  0.4  &  2.5  &  11.354  &    17 &  \\
2864        &  6.0  &  77.3   &  68.2   & 0.150  &  0.8   &  0.0  &  1.0  &  11.649  &    24 &  \\
2910        &  4.8  &  67.4   &  15.5   & 0.425  &  0.1   &  0.0  &  0.4  &  10.913  &     5 &  \\
3179\tablenotemark{a} &  6.5  &  85.1   &  85.1   & 0.000  &  0.5   &  0.0  &  1.0  &  11.667  &    13 & Mobasher et al. (2005) \\
3348        &  5.0  &  66.6   &  32.9   & 0.025  &  0.7   &  0.1  &  0.2  &  11.543  &    10 & X--ray detected \\
3361        &  5.1  &  99.6   &  51.3   & 0.000  &  0.8   &  0.3  &  1.0  &  10.756  &    12 &  \\
4071        &  5.1  &  62.4   &  51.2   & 0.250  &  0.2   &  0.0  &  1.0  &  11.121  &     6 &  \\
4135        &  4.7  &  70.8   &  34.6   & 0.375  &  0.2   &  0.8  &  0.4  &  10.924  &     6 &  \\
4550        &  4.8  &  79.4   &  22.2   & 0.150  &  1.0   &  0.2  &  2.5  &  11.239  &    16 &  \\
5197        &  5.2  &  99.5   &  93.2   & 0.000  &  0.9   &  0.3  &  0.2  &  10.896  &    17 & Spectroscopically confirmed \\
\enddata
\tablenotetext{a}{The JD2 (BBG\#3179) results are based on Monte Carlo simulation using photometric data from the Hubble UDF
(see Sect.~\ref{individual}; Mobasher et al. 2005)}
\end{deluxetable}

\clearpage
\begin{deluxetable}{lccccccc}
\tablewidth{0pt}
\tabletypesize{\scriptsize}

\tablecaption{Comparison of spectroscopic and photometric redshifts
for sources with $z_{\mathrm{spec}} > 4$\label{comparison}}

\tablehead{
\multicolumn{1}{c}{Galaxy}                                &
\multicolumn{1}{c}{Redshift}                              &
\multicolumn{1}{c}{E$_{\mathrm{B-V}}$}                    &
\multicolumn{1}{c}{age}                                   &
\multicolumn{1}{c}{$\tau$}                                &
\multicolumn{1}{c}{Z}                                     &
\multicolumn{1}{c}{Log(M$_{*}$)}                          &
\multicolumn{1}{c}{Note}                                  \\
\multicolumn{1}{c}{}                                     &
\multicolumn{1}{c}{$z_{\mathrm{phot}}$} &
\multicolumn{1}{c}{$z_{\mathrm{spec}}$}  &
\multicolumn{1}{c}{}                                      &
\multicolumn{1}{c}{Gyr}                                   &
\multicolumn{1}{c}{Z$_{\odot}$}                           &
\multicolumn{1}{c}{M$_{\odot}$}                           &
\multicolumn{1}{c}{}                                      
}
\startdata
Yan \#1       &  5.7   & 0.0 & 1.0 & 1.0 & 0.2 & 10.29 & Our result \\
                     &  5.83 & 0.0 & 0.5 & 0.0 & 1.0 & 10.53 & Yan et al. (2005) \\
%\hline
Yan \#4       & 4.8    & 0.025 & 0.1 & 0.2 & 0.2 &  8.99 & Our result \\
                     & 5.05 & ---        & ---   & ---   & ---  & ---      &  Yan et al. (2005) \\
%\hline
Yan \#5       &  5.8   &  0.0     & 1.0 & 0.4 & 0.2 & 10.35 & Our result \\
                     &  5.90 & 0.0      & 0.9 & 0.2 & 1.0 & 10.58 & Yan et al. (2005) \\
%\hline
Yan \#6       &  4.5   &  0.0     & 0.4 & 0.1 & 0.2 &   9.91 & Our result \\
                     &  4.65 & 0.0      & 1.3 & 0.4 & 0.01 & 10.34 & Yan et al. (2005) \\
%\hline
Yan \#7       &  4.5   &  0.0     & 0.7 & 1.0 & 0.2 &   9.81 & Our result \\
                     &  4.78 & ---        & ---    & ---   & ---   & ---        & Yan et al. (2005) \\
%\hline
Yan \#15     &  4.8   &  0.10   & 0.035 & 0.0 & 0.2 &  9.58 & Our result \\
                     &  5.49 & 0.0      & 1.0 & 0.6 & 1.0 & 10.34 & Yan et al. (2005) \\
%\hline
SBM03\#3  &  5.7   & 0.0     & 0.5 & 1.0 & 0.2 & 10.11 & Our result \\
                     &  5.78 & 0.0     & 0.6 & 0.5 & 1.0 & 10.68 & Eyles et al. (2005) \\
%\hline
GLARE\#3001 &  5.7   &  0.175 & 0.010 & 0.0 & 1.0 &   9.32 & Our result \\
                           &  5.79 & ---        & ---        & ---  & ---   & ---       & Eyles et al. (2005) \\
%\hline
HCM\,6A          &  6.6   &  0.0  &  0.3             &  1.0 & 0.2 &   9.6         &  Our result\tablenotemark{(1)}  \\
                          &  6.56 &  0.2  &  0.2$-$0.6 & ---    & ---   & 9.0$-$9.9  & Hu et al. (2002); Chary et al. (2005) \\
%\hline
35\_4142         & 4.5    & 0.3  & 0.005 & 0.6 & 0.4 &  9.39 & Our result \\
                          & 4.91  & 0.0  & 0.161 & csf  & 1.0 &  9.34 & Vanzella et al. (2006); Stark et al. (2006)  \\
%\hline
35\_6626         & 5.4    & 0.05  & 0.015 & 0.0    & 1.0 &  8.87 & Our result \\
                          & 5.25  & 0.0  & 0.143   & 0.07  & 1.0 &  9.32 & Vanzella et al. (2006); Stark et al. (2006)\tablenotemark{(3)} \\
%\hline
35\_6867         & 0.8    & 0.3   & 0.2        & 0.0    & 2.5 &  9.02   & Our result \\
                          & 4.42  & 0.01 & 0.360   & 0.10  & 1.0 & 10.37 & Vanzella et al. (2006); Stark et al. (2006) \\
%\hline
32\_8020         & 5.2    & 0.075 & 1.0        & 0.6    & 0.4 & 11.00   & Our result (BBG\#5197)\\
                          & 5.55  & 0.0      & 0.905   & 0.30  & 1.0 & 11.16   & Vanzella et al. (2006); Stark et al. (2006) \\
%\hline
35\_9350         & 5.3    & 0.0   & 0.1        & 0.2    & 0.4 &  9.16   & Our result \\
                          & 5.28  & 0.00 & 0.255   & csf     & 1.0 &  9.33 & Vanzella et al. (2006); Stark et al. (2006) \\
%\hline
34\_9738         & 0.8    & 0.15 & 0.5        & 0.0    & 2.5 &  8.75   & Our result \\
                          & 4.79  & 0.00 & 0.360   & 0.10  & 1.0 & 10.13 & Vanzella et al. (2006); Stark et al. (2006) \\
%\hline
32\_10232      & 0.9    & 0.1   & 0.6        & 0.1    & 2.5 &  8.80   & Our result \\
                          & 4.90  & 0.01 & 0.255   & 0.07 & 1.0 & 10.06  & Vanzella et al. (2006); Stark et al. (2006) \\
%\hline
33\_10340      & 4.6    & 0.1    & 0.2        & 0.4    & 0.4 & 10.06   & Our result ($z \approx 4$ BBG)\\
                          & 4.44  & 0.24 & 0.018   & 0.10  & 1.0 & 11.28   & Vanzella et al. (2006); Stark et al. (2006) \\
%\hline
35\_11280      & 4.7    & 0.0   & 0.8        & 0.3    & 0.4 &  9.82   & Our result \\
                          & 4.99  & ---    & ---          & ---     & ---   & ---        & Stark et al. (2006); Stark et al. (2006) \\
%\hline
35\_14097      & 4.6    & 0.175 & 0.07      & 0.4    & 0.8 &  9.44   & Our result \\
                          & 4.60 & 0.05   & 0.255    & 0.20  & 1.0 &  9.33   & Vanzella et al. (2006); Stark et al. (2006) \\
%\hline
31\_14602      & 4.2    & 0.025 & 0.3        & 0.0    & 0.4 & 10.70   & Our result \\
                          & 4.76  & 0.00  & 1.000    & 0.30  & 1.0 & 11.10 & Vanzella et al. (2006); Stark et al. (2006) \\
%\hline
21\_23040      & 5.0    & 0.0   & 0.6        & 0.2    & 1.0 & 10.28   & Our result \\
                          & 4.40 & 0.53 & 0.001   & 0.00  & 1.0 &  8.43    & Vanzella et al. (2006); Stark et al. (2006) \\
%\hline
23\_23051      & 4.9    & 0.0   & 0.4        & 0.2    & 0.4 &  9.75   & Our result \\
                          & 4.84  & 0.00 & 0.286   & 0.10  & 1.0 &  9.86  & Vanzella et al. (2006); Stark et al. (2006) \\
%\hline
21\_24396      & 5.3    & 0.175 & 0.01      & 0.0    & 1.0 &  9.57   & Our result \\
                          & 5.37  & 0.17   & 0.009   & 0.00  & 1.0 &  8.40   & Vanzella et al. (2006); Stark et al. (2006) \\
%\hline
22\_25323       & 4.9    & 0.0   & 0.4        & 0.2    & 2.5 &  9.79   & Our result \\
                          & 4.76  & 0.32 & 0.003   & 0.00  & 1.0 &  8.43   & Vanzella et al. (2006); Stark et al. (2006) \\
%\hline
33\_10388\tablenotemark{(2)}  & 4.6    & 0.0   & 0.7        & 0.2    & 0.8 & 10.63   & Our result ($z \approx 4$ BBG)\\
                                    & 4.50  & ---     & ---          & ---      & ---  & ----        & Vanzella et al. (2006); Stark et al. (2006)
%\hline
\enddata
\tablenotetext{(1)}{Stellar mass is corrected for magnification due to lensing by a factor of 4.5}
\tablenotetext{(2)}{Uncertain spectroscopic redshift (Vanzella et al. 2006)}
\tablenotetext{(3)}{The id's are from Stark et al. (2006), redshifts from Vanzella et al. (2006)}
\end{deluxetable}

\clearpage

\begin{deluxetable}{llc}
\tablewidth{0pt}
\tabletypesize{\scriptsize}

\tablecaption{Stellar mass densities\tablenotemark{a}\label{stellarmass}}

\tablehead{
\multicolumn{1}{c}{Redshift}                                                                    &
\multicolumn{1}{c}{$\log(\rho_{*} / \mathrm{M}_{\odot}\,\mathrm{Mpc}^{-3})$}  &
\multicolumn{1}{c}{Reference}                                                                                               
}
\startdata
0.0              & $8.60$                  & Bell et al. (2003)       \\
\ \\
0.1              & $8.59^{+0.04}_{-0.04}$  & Rudnick et al. (2006)    \\
\ \\
$0.5-1.4$        & $8.46^{+0.07}_{-0.07}$  & Dickinson et al. (2003)  \\
\ \\
2.0              & $7.48^{+0.12}_{-0.16}$  & Rudnick et al. (2003)    \\
\ \\
$2.0-2.5$        & $7.58^{+0.11}_{-0.07}$  & Dickinson et al. (2003)  \\
\ \\
2.5              & $7.60^{+0.04}_{-0.04}$  & Fontana et al. (2006)    \\
\ \\
$2.5-3.0$        & $7.52^{+0.23}_{-0.14}$  & Dickinson et al. (2003)  \\
\ \\
2.8              & $7.59^{+0.10}_{-0.10}$  & Rudnick et al. (2006)    \\
\ \\
3.5              & $7.23^{+0.12}_{-0.12}$  & Fontana et al. (2006)    \\
\ \\
$4.5$            & $7.60^{+0.15}_{-0.25}$  & Drory et al. (2005)      \\
\ \\
5.0              & $>6.78$                 & Stark et al. (2006)      \\
\ \\
$5.2 \pm 0.5$    & $>6.90$ ($>6.15)$\tablenotemark{b} & This paper               \\
\ \\ 
6.0              & $>(6.04 - 6.83)$        & Yan et al. (2006)        \\   
\ \\
\hline        
\enddata
\tablenotetext{a}{Not a complete sample of mass density estimates (cf. Rudnick et al. 2006
for further references)}
\tablenotetext{b}{no--MIPS sample}
\end{deluxetable}

\clearpage

\clearpage

\clearpage

\clearpage

\clearpage

\clearpage

\clearpage

\clearpage

\clearpage

\clearpage

\clearpage

\clearpage

\clearpage

\clearpage

\clearpage

\clearpage

\clearpage

\clearpage

\clearpage

\clearpage

\end{document}